\documentclass{aa}
\usepackage[varg]{txfonts}
\usepackage{natbib}
\usepackage{hyperref}
\bibpunct{(}{)}{;}{a}{}{,} 
\usepackage{subfig}
\usepackage{graphicx}

\hypersetup{colorlinks=true,urlcolor=cyan}

\providecommand{\Tprime}{\ensuremath{T_\textrm{A}^\prime}}
\providecommand{\Tstar}{\ensuremath{T_\textrm{A}^*}}
\providecommand{\Tsys}{\ensuremath{T_{sys}}}
\providecommand{\HIFIHB}{\it{HB} \normalfont}
\providecommand{\HDRG}{\it{HDRG} \normalfont}
\providecommand{\HPSD}{\it{HPS} \normalfont}
\providecommand{\HPDP}{{\it HPDP}}
\providecommand{\Herschel}{{\it Herschel} }

\begin{document}

\title{Data processing pipeline for {\it Herschel} HIFI \thanks{\it{Herschel} \rm was an ESA space 
observatory with science instruments provided by European-led Principal Investigator consortia and 
with important participation from NASA.}}

\author{R.F. Shipman \inst{1,2} 
        \and S. F. Beaulieu \inst{3}
        \and D. Teyssier \inst{4,5}
        \and P. Morris \inst{6}
    	 \and M. Rengel \inst{4,7}	 
	 \and C. McCoey \inst{3}              
	 \and K. Edwards \inst{3}  
        \and D. Kester \inst{1}
         \and A. Lorenzani \inst{8}   
         \and O. Coeur-Joly \inst{9}
         \and M. Melchior \inst{10}   
         \and J. Xie \inst{5}       	
         \and E. Sanchez \inst{14}
  	\and P. Zaal \inst{1}
         \and I. Avruch \inst{1}
         \and C. Borys \inst{6}
         \and J. Braine \inst{11}
         \and C. Comito \inst{13}
         \and B. Delforge \inst{2}
         \and F. Herpin \inst{11}
         \and A. Hoac \dag  \inst{6} 
	\and W. Kwon \inst{1,16}
	\and S. D. Lord \inst{6}
	\and A. Marston \inst{15}
         \and M. Mueller \inst{1,2}
	\and M. Olberg \inst{12}	
	\and V. Ossenkopf \inst{13}
	\and E. Puga \inst{4}
	\and M. Akyilmaz-Yabaci \inst{13}
	  }

\institute{SRON Netherlands Institute for Space Research, Netherlands
      \and Kapteyn Astronomical Institute, University of Groningen, Netherlands
      \and University of Waterloo, Waterloo, Canada
      \and Herschel Science Center, ESAC, Villefranca, Spain
      \and Telespazio Vega UK Ltd for ESA, European Space Astronomy Centre (ESA/ESAC), Operations Department, Villanueva de la Cañada (Madrid), Spain
      \and IPAC, Mail Code 100-22, Caltech, 1200 E. California Blvd., Pasadena, CA 91125 USA 
      \and Max Planck Institute f\"{u}r Sonnensystemforschung, G\"{o}ttingen, Germany
      \and Osservatorio Astrofisico di Arcetri, Florence, Italy
      \and IRAP, Institut de Recherche en Astrophysique et Planetologie CNRS/UPS, Toulouse, France
      \and FachHochschule Nordwest, Switzerland
      \and Laboratoire d'astrophysique de Bordeaux, Univ. Bordeaux, CNRS, B18N, all\'ee Geoffroy Saint-Hilaire, 33615 Pessac, France
      \and Onsala Space Observatory, Onsala, Sweden
      \and University of Cologne, Cologne, Germany
      \and ESTEC, Noordwijk, The Netherlands
      \and ESA, The Space Telescope Science Institute, Baltimore, USA
      \and Korea Astronomy and Space Science Institute (KASI), 776 Daedeokdae-ro, Yuseong-gu, Daejeon 34055, Republic of Korea    
      }

\date{}

\abstract {The HIFI instrument on the {\it Herschel Space Observatory} performed over 9100 astronomical 
observations, almost 900 of which were calibration observations in the course of the nearly four-year \Herschel 
mission. The data from each observation had to be converted from raw telemetry into calibrated products 
and were included in the Herschel Science Archive.}
{
The HIFI pipeline was designed to provide robust conversion from raw telemetry into calibrated data throughout 
all phases of the HIFI missions.  Pre-launch laboratory testing was supported as were routine mission operations.
}
{A modular software design allowed components to be easily added, removed, amended and/or extended as the 
understanding of the HIFI data developed during and after mission operations.  
}
{The HIFI pipeline processed data from all HIFI observing modes within the \Herschel automated processing 
environment as well as within an interactive environment. The same software can be used by the general 
astronomical community to reprocess any standard HIFI observation. The pipeline also recorded the consistency 
of processing results and provided automated quality reports. Many pipeline modules were in use since the 
HIFI pre-launch instrument level testing.
}
{Processing in steps facilitated data analysis to discover and address instrument artefacts and uncertainties. 
The availability of the same pipeline components from pre-launch throughout the mission made for well-understood, 
tested, and stable processing. A smooth transition from one phase to the next significantly enhanced processing 
reliability and robustness. 
}

\keywords{Instrumentation: spectrographs-- Methods: data analysis}

\maketitle
{\let\thefootnote\relax\footnotetext{\dag  Deceased September 2016}}
\setcounter{footnote}{0}

\section{Introduction to the HIFI instrument}
\label{sec:intro}

The Heterodyne Instrument for the Far Infrared \citep[HIFI;][]{degraauw2010} was a heterodyne spectrometer 
that operated on board the Herschel Space Observatory \citep{pilbratt2010} between May 2009 and April 
2013. 

In heterodyne spectroscopy, the incident sky signal at frequency $\nu_{sky}$ is mixed with a locally generated 
signal (Local Oscillator, LO) at $\nu_\textrm{{LO}}$ close to the sky frequency and passed onto a non-linear 
receiver. The non-linear receiver is sensitive to the frequency "beats" that the mixing produces. The beat frequencies 
($||\nu_{sky}-\nu_{\textrm{LO}}||$ ) cover a band of frequencies at a significantly lower frequency that can be 
readily analysed. A frequency within this band is called the intermediate frequency (IF) of the instrument. Although 
the IF is at a much lower frequency than the sky it retains the information of the sky signal. The IF band is then 
amplified and passed on to spectrometers. The HIFI receivers were dual sideband meaning that information 
at any $\nu_{\textrm{IF}}$ was the combination of information at $\nu_{\textrm{LO}} +\nu_{\textrm{IF}}$ (upper 
sideband USB) and $\nu_{\textrm{LO}} -\nu_{\textrm{IF}}$ (lower sideband LSB). 

HIFI was made up of 14 dual-sideband (DSB) receivers covering frequencies from 480 GHz to 1906 GHz in two 
polarisations (horizontal (H) and vertical (V)). The LO produced continuous signal from 480 to 1272 GHz in five 
continuous frequency bands (bands 1 to 5) and two high frequency bands between 1430 and 1906 GHz (bands 
6 and 7). The mixers which combined the sky signal and the LO signal were superconductor-insulator-superconductor 
(SIS) mixers for bands 1 to 5 and hot electron bolometer (HEB) mixers for bands 6 and 7. The IFs for receiver 
bands 1 to 5 covered frequencies from 4 to 8 GHz, while the IFs for bands 6 and 7 covered frequencies from 2.4 
to 4.8 GHz.

HIFI was equiped with two spectrometer backends, the acousto-optical Wide Band Spectrometer (WBS), and 
the High Resolution Spectrometer (HRS) autocorrelator. For intensity calibration purposes, HIFI used internal 
hot and cold loads, and a chopping mechanism to choose between the loads and the sky-on target, and 
sky-3\arcmin\ off target. Extensive online HIFI documentation can be found in the Herschel Explanatory Legacy Library
\footnote{\href{https://www.cosmos.esa.int/web/herschel/legacy-documentation-hifi}{https://www.cosmos.esa.int/web/herschel/legacy-documentation-hifi}\label{hell_fn}}
where a thorough overview of the HIFI instrument is documented in the HIFI Handbook \citep[hereafter \HIFIHB]{HIFIH}.

The conversion of the raw HIFI telemetry to scientifically usable products was performed by the HIFI pipeline. 
The HIFI pipeline and all associated data products were developed within the Herschel Common Software System 
(HCSS). The goal of the HCSS was to  bring the three \Herschel instruments together with the satellite into one 
common software framework. This framework had broad implications ranging from defining the underlining software 
language (Java) and databases to implementing common product definitions for images, spectra, and data cubes. 
It is within the HCSS that all instrument pipelines operated within the Standard Product Generation (SPG) component. 
Within this context, HIFI data products begin very HIFI specific and become more HIFI independent at the highest 
level of pipeline processing.

The pipeline was designed by the HIFI Instrument Control Centre (ICC) to process raw telemetry for the entire 
{\it Herschel} HIFI mission from the  pre-launch laboratory phase  to the post mission archive phase, a roughly 15 
year period. To accommodate the many anticipated and unanticipated changes inherent in designing software over 
such a large time span, the pipeline was divided into processing levels comprised of (mostly) independent processing 
steps.  

The HIFI pipeline produced data products for all HIFI observing modes as well as sensible processing for HIFI 
engineering modes (not discussed in this article). The pipeline also operated in both an interactive user environment 
for the calibration scientists and instrument engineers as well as within a completely hands-off environment of bulk 
processing (SPG).

During the mission, the HIFI instrument performed over 9100 observations where about 10\% were calibration
observations. The pipeline processed data from these observations are publicly available from the Herschel Science 
Archive\footnote{\href{http://www.cosmos.esa.int/web/herschel/science-archive}{http://www.cosmos.esa.int/web/herschel/science-archive}\label{hsa_fn}} (HSA). 

The purpose of this article is to describe the HIFI  standard processing pipeline for HIFI astronomical observations. 
This article focuses on the issues of the pipeline that mainly impact a general astronomer or archive user. A more 
detailed paper describing the implementation of the pipeline software will be presented in 
\citep[hereafter Paper II]{edwards2017}.

This article is organised as follows. Section \ref{sec:modes} reviews the observing modes which were available 
for the HIFI instrument during nominal operations. The relevant data processing concepts are described in 
section \ref{sec:concepts}. In section \ref{sec:levels}, selected pipeline steps are described. Section 
\ref{sec:quality} reviews the various quality assessment steps that are taken within the pipeline. Section 
\ref{sec:additional} identifies a number of further processing steps beyond the pipeline which might be
useful for full exploitation of the data. Section \ref{sec:discussion} reviews some of the main lessons learned 
in the process of developing a data processing pipeline. We end with a summary in Section \ref{sec:summary}.

\section{Observing modes}
\label{sec:modes}

The HIFI pipeline is the implementation of the HIFI instrument calibration framework (flux calibration 
\citep{ossenkopf2003} and frequency calibration \citep{herpin2003}) and brings together instrument, 
satellite, and observation information that has bearing on the observational results. The main principle 
of observing with HIFI was to quickly switch between the astronomical source and some reference signal 
to correct for instabilities that were due to drift within the signal chain. To correct for standing waves, observations 
of a blank sky (OFF) were performed. The OFF measurements required nodding the telescope and were 
taken at a slower rate. For the faster referencing, HIFI made use of internal sources (hot and cold loads), 
modulation of the LO, or two chopped positions on (assumed) blank sky, each at a fixed distance of 
$3\arcmin$ on either side of the source. The quick modulation measurements corrected for the spectral 
response of the system (bandpass), whereas the OFF measurement corrected for instrument response 
drift and standing waves.

The timing and frequency of the referencing are governed by the Allan instability time \citep{2008ossenkopf} 
and the desired observation signal-to-noise ratio \citep{morris&ossenkopf2015}. Essentially, the timing within an 
observation was optimised to make efficient observations in the radiometric regime of the mixer. 

The calibration reference schemes available to astronomers through the Herschel Observation Planning 
Tool (HSpot) \citep{HSPOTM} are listed below.

\begin{itemize}
\item Position Switch (PSW): this modulated the signal against an OFF position by slewing the telescope. 
\item Dual Beam Switch (DBS): this used the internal chopper to modulate the input signal to an OFF position 
$3\arcmin$ from the source,  then slewed the satellite $3\arcmin$ to the other side of the source (nodding), 
and then it chopped back to the ON. 
\item Frequency Switch (FSW): this modulated the LO to a frequency ($\nu_0 + \nu_{throw}$, where $\nu_{throw}$ 
can be positive or negative). The correction for instrument response drift was achieved by performing an FSW 
on an OFF position. 
\item Load Chop: this used the internal chopper to modulate the signal against the internal cold load. The correction 
for instrument response drift was achieved by performing a load chop on an OFF position. 
 \end{itemize} 

HIFI observed using astronomical observing templates (AOT) for different referencing schemes for the following 
three different modes: single spectra, spectral mapping, and spectral scans. These modes and their allowed 
referencing schemes are described below.

\begin{itemize}
\item{Single Point mode: \label{sec:modes:point}}
spectra taken at a single position on the sky in one LO setting were called "Point" spectra (sometimes referred 
to as single-point spectra). Point spectra could be observed using the referencing schemes described above.  

\item{Spectral Mapping mode: \label{sec:modes:map}} 
spectra taken at multiple positions at the same LO resulted in spectral maps. There were two main spectral 
mapping modes, raster and on-the-fly (OTF). The raster mode always used the DBS referencing scheme. 
In the OTF, the \Herschel satellite scanned a pattern while HIFI integrated on the sky using one of the reference 
schemes, FWS, Load-Chop, or PSW. 

\item{Spectral Scan mode: \label{sec:modes:scan}}
spectral surveys were obtained at a fixed position on the sky while  scanning  a wide LO range within a 
given mixer band and were performed in DBS, FSW, or Load-Chop referencing schemes. Position-switch 
referencing scheme was not available for this mode.
\end{itemize}

Table \ref{tbl:modes} lists the available AOTs (mode and referencing scheme) that were available for any 
given HIFI observation, and lists the mode names as they are found in archived HIFI data products. For 
historical reasons, the modes are identified as \texttt{CUSMODE} within the HIFI data products. A detailed 
description of each mode and referencing scheme can be found in \citet{morris&ossenkopf2015}.

Observations are comprised of the source and OFF position, the frequency, the integration time (or desired 
noise), and the AOT. Together, these constituted an Astronomical Observing Request (AOR). Every scientific 
observation in the HSA is the result of an AOR. The HSA also contains calibration observations. Usually, 
these observations made use of standard observing templates, but they could have been taken with some 
different observing parameters. However, for automatic processing the fundamentals of these observations 
did not differ significantly from a standard AOT. 

\begin{table*}
\begin{center}
\caption{Allowed CUSMODE/INSTMODE  keywords for AOTs \label{tbl:modes}}
\begin{tabular}{rrrrr}
\hline \hline
AOT:  CUSMODE name & Mode & Referencing Scheme & OFF & Remarks \\
\hline
HifiPointModeDBS & Single Point mode & Dual Beam Switch & 3\arcmin & \\
HifiPointModeFastDBS & Single Point mode & Dual Beam Switch & 3\arcmin & timing variant of DBS \\
HifiPointModePositionSwitch & Single Point mode & Position Switch & yes & \\
HifiPointModeFSwitch & Single Point mode & Frequency Switch & yes & \\
HifiPointModeFSwitchNoRef &Single Point mode & Frequency Switch & no & \\
HifiPointModeLoadChop & Single Point mode & Load Chop & yes & \\
HifiPointModeLoadChopNoRef & Single Point mode& Load Chop & no & \\
HifiMappingModeDBSRaster & Spectral Mapping mode & Dual Beam Switch & 3\arcmin & \\
HifiMappingModeFastDBSRaster & Spectral Mapping mode & Dual Beam Switch & 3\arcmin & timing variant of DBS \\
HifiMappingModeOTF & Spectral Mapping mode & Position Switch & yes & \\
HifiMappingModeFSwitchOTF & Spectral Mapping mode& Frequency Switch & yes & \\
HifiMappingModeFSwitchOTFNoRef & Spectral Mapping mode& Frequency Switch & no & \\ 
HifiMappingModeLoadChopOTF & Spectral Mapping mode & Load Chop & yes & \\
HifiMappingModeLoadChopOTFNoRef & Spectral Mapping mode & Load Chop & no & \\
HifiSScanModeDBS & Spectral Scan mode & Dual Beam Switch & 3\arcmin & \\
HifiSScanModeFastDBS & Spectral Scan mode & Dual Beam Switch & 3\arcmin & timing variant of DBS \\
HifiSScanModeFSwitch & Spectral Scan mode & Frequency Switch & yes & \\
HifiSScanModeFSwitchNoRef &Spectral Scan mode& Frequency Switch & no & \\
\hline
\end{tabular}
\end{center}
\end{table*}

section{Concepts of HIFI pipeline processing}
\label{sec:concepts}
\subsection{One pipeline for all observations}
\label{sec:concepts:one}
The HIFI pipeline is the component of the SPG \citep{ott2010} that processed HIFI-specific data to populate 
the HSA. The pipeline was specifically designed for processing data on a polarisation basis from each of the 
spectrometers used in the observation. This can include low-level diagnostic housekeeping data even for a 
spectrometer that was not employed for science. The HIFI data were automatically processed through the 
pipeline and included in the HSA. All data were reprocessed during periodic bulk reprocessing after major 
software updates. The last bulk reprocessing for HIFI was completed with SPG version 14.1. About a dozen 
observations, however, had to be processed with SPG version 14.2 to fix an isolated issue.

Processing HIFI data makes best use of the fact that the same processing steps are needed regardless 
of observing mode and referencing scheme or spectrometer. For example, after the calibration of the different 
spectrometers, all processing steps are the same for either the HRS or the WBS data. The following sections 
{\bf contain} pointers that describe where differences in the pipeline exist.

\subsection{Data format}
\label{sec:concepts:format}
In the pre-launch phase, the main use of the pipeline was to process and test the spectrometer backend data. 
A HIFI data block consisted of the flux and frequency of all channels in a single readout of one of the backend 
spectrometers. An integration was  built up of multiple readouts. Given the potential for instrument drifts, the 
readouts were not immediately co-added. The frequency values were not assumed to be fixed to a channel. 
At any one time, an instrument readout consisted of fluxes and frequencies at each channel of the spectrometer. 

HIFI took observations as a function of time steps as prescribed by the Observing Mode command sequence that was
generated to carry out the astronomer's AOR. At any moment, HIFI was executing the flow of the observation 
template, be it configuring the instrument, slewing the satellite, integrating on source or off source, or an internal 
load. At any one time, HIFI observed intensity as a function of frequency. When HIFI was performing the same 
sort of repeated integration, a two-dimensional spectrum was being built over time. The container of such a 
spectrum was called a \texttt{HifiSpectrumDataset}. A given observation sequence could have hundreds or 
thousands of individual integrations. Instead of creating hundreds of thousands of individual spectra, the 
spectra were grouped in time sequences that formed 2D blocks of spectra. Each block belonging 
to a given activity as defined by the AOT. These blocks of spectra were placed into a \texttt{HifiTimelineProduct}. 
All HIFI processing was made on these 2D spectra. Listing spectra as 2D entities is a non-standard format that 
was not directly usable in most software packages for spectra. At the very end of the pipeline when all the 
observing mode specific time sequences of spectra were reduced, the resulting spectra were reformatted 
to a standard data format for spectra (frequency vs. 1D flux). Recently, GILDAS/CLASS 
\citep{bardeau2015a, bardeau2015b} has been updated to read HIFI data directly in its native format.

All HIFI data come from the HSA in the form of a directory-tree structure. The tree structure is a Java 
HCSS-specific set of product bindings and pointers to other products, objects, and metadata. When inspected 
in the Herschel Interactive Processing Environment (HIPE) \citep{ott2010}, main branches are referred to as 
 context. A context is a special type of product that links other products in a coherent description and can 
be thought of as an inventory or catalogue of products. The HIFI-processed observation consists of many 
such contexts enclosed within one observation context. The observation context is therefore a 
product container that is comprised of various layers of science data, auxiliary data, calibration data, and 
other information about the observation.

\texttt{MetaData} form another important data component and describe each data product in more detail. 
\texttt{MetaData} contain parameters such as the source name, observation id, or the LO frequency, but 
also observation performance parameters. The \texttt{MetaData} perform the same function as \texttt{FITS} 
headers but for all products and sub-products. More detailed information on any HIFI data product is found in {\it HIFI Products Explained}\citep{hpe2017}\footnote{\href{http://herschel.esac.esa.int/twiki/pub/Public/HifiDocsEditableTable/HIFI\_Products\_Explained\_v1.pdf}    {HIFI\_Products\_Explained\_v1.pdf in HIFI online documentation}}\label{hpe_fn}.

\subsection{Calibration}
\label{sec:concepts:calibration}
Calibrating HIFI data involved converting raw instrument outputs into scientifically usable quantities, that is, fluxes 
(antenna temperatures) and frequencies. It also entailed assessing the performance of the calibration steps, that is, 
whether were the resulting parameters where within the expected ranges. This last item required information from the AOR that 
defined the observation. To accommodate these concepts, the HIFI calibration was separated into three main
sections: 
\begin{itemize}
\item Downlink: these are pipeline input parameters that were measured independently of any one observation 
and were relatively constant over a span of time. Some example parameters are: load coupling coefficients and sideband 
ratios.
\item Pipeline-out: these were pipeline-calculated parameters that were the results of pipeline steps, for example, the 
system noise temperature (\Tsys).
\item Uplink: these were parameters from the AOR that specified how the observation was to be performed. The noise goals were specified in the uplink calibration.  
\end{itemize}

When obtaining the \texttt{Calibration} product for an observation in the HSA, the retrieved data will reflect 
these three calibration areas and present the data either as \texttt{FITS} binary tables or spectra within each. 
A full description of the HIFI calibration tree is given in the Chapter 5.6 of the \HIFIHB.

Two especially notable products in the \texttt{pipeline-out} calibration are the OFF spectra and the uncertainty 
table. These are described in more detail below.

\subsubsection{OFF spectra}
\label{sec:offspectra}
For each observation with a dedicated reference position observation, also called "OFF spectra", the resulting 
OFF spectra were calculated and attached to the \texttt{pipeline-out} calibration branch under the directory 
\texttt{ReferenceSpectra}. The OFF spectra are in the native HIFI FITS format (Section \ref{sec:concepts:format}).  

Since these spectra are generally useful to identify potential emission contamination in the blank sky 
(OFF position), the spectra are available as Highly Processed Data Products (\HPDP) 
\footnote{\href{http://www.cosmos.esa.int/web/herschel/highly-processed-data-products}{http://www.cosmos.esa.int/web/herschel/highly-processed-data-products}\label{hpdp_fn}} 
from the HSA or via a dedicated link from the Herschel Science Centre. The \HPDP ~data-set offers the additional 
advantage of providing OFF spectra as cubes for DBS raster maps and deconvolved spectra for spectral scans, 
which are more directly comparable to their ON-source counterparts.

For DBS referencing, the OFF spectra are made from the difference of the two sky positions on either side of the 
source, whereas for other referencing schemes, the OFF spectra are the spectra of the OFF position. The 
position-switch referencing scheme uses the OFF position for the fast referencing, and the OFF observation itself 
does not have an independent reference. In this case, a referencing scheme for the OFF is made from the average 
of the cold-load data for that observation. Without the benefit of a double difference, the OFF spectra can display 
significant standing waves which can be addressed through interactive processing (see Section \ref{extra:fhf}). 

\subsubsection{Uncertainty tables}
\label{sec:uncertainty}
The HIFI flux calibration uncertainty is broken down between the various components that enter the general 
calibration equation \citep{ossenkopf2015}. The HIFI pipeline implements the propagation of errors  and 
provides uncertainty tables per polarisation and sideband for each observation. For each uncertainty 
component, the uncertainty is given as coefficients of a polynomial describing the possible dependency on 
the Intermediate Frequency (IF). We note that the uncertainties in the table apply directly to the Level 2 or Level 
2.5 products calibrated on the \Tstar scale.

For spectral maps in particular, only one uncertainty budget table is provided per spectrometer, polarisation, 
and sideband because no distinction is made between the respective map points.

The uncertainty table of a given observation can be downloaded from the HSA by retrieving the \texttt{Calibration} 
product. The FITS table is found under the directory \texttt{pipeline-out} : \texttt{Uncertainty} : for instance for WBS.  
Similarly to the OFF spectra dataset, the uncertainty tables can also be retrieved as a stand-alone \HPDP ~
from the HSA, or at the link mentioned in the previous section.

\section{Processing levels \label{sec:levels}}
The operation of HIFI was performed in predefined command blocks set by the AOT command logic built for 
the astronomer's AOR. Some command blocks performed standard instrument set-up steps where the processing 
is well known and the results can be checked for consistency. However, some of the command blocks  were
designed to integrate on an internal reference or on the OFF position with the purpose of correcting instrument 
instabilities. The pipeline processed these command blocks as prescribed by the AOR assuming a "best case" 
scenario -- best case being that the timing of references, OFFs, and all calibration activities were sufficient to 
correct the HIFI data. For a review of the performance of the AOTs see \citet{morris&ossenkopf2015}. In light 
of the potential for corrections to have been incomplete, the pipeline processed data up to reasonable stopping 
points, {\bf referred to as pipeline levels}. These pipeline levels are described below. A series of pipeline flow diagrams are found 
in {Figures \ref{fig:level0} to \ref{fig:level2}} for each level. A detailed description of all the pipeline steps can be 
found in the HIFI Pipeline Specification\footnote{\href{http://herschel.esac.esa.int/hcss-doc-15.0/load/hifi_pipeline/html/hifi_pipeline.html}{http://herschel.esac.esa.int/hcss-doc-15.0/load/hifi\_pipeline/html/hifi\_pipeline.html}\label{hpsm_fn}} (\HPSD). 
Here, a brief description of each level is presented:

\subsection{Level 0 \label{sec:levels:level0}}
Level 0 was the rawest form of HIFI-processed data available for inspection. It has been minimally manipulated 
to create the initial {\it HifiTimelineProduct} (HTP). These data contained all the readouts of the HIFI spectrometers 
plus satellite-pointing information associated with them, and have undergone several "sanity checks" to flag 
any incidences of "out of limit" housekeeping parameters.  
  
Figure~\ref{fig:level0} shows the flow of the Level 0 pipeline. The Level 0 pipeline created an HTP for each 
spectrometer used in the observation, which in turn contained a data set for each integration type in the
observation. The chopper state was used at this stage to help identify and verify the integration type.

Level 0 contained many more data blocks than higher levels of processing. Integration types reported the 
activity that was executed, be it an internal load measurement or observing the reference position on the 
sky. The data types were recorded in building block ids, and these are listed in Table \ref{tab:bbids} and online
in the section "Standard Observing Modes" of the {\HPSD}. The table indicates whether the data block resulted in 
science data or was a spectrometer calibration step and to which backend that step belonged. The column 
"line" specifies a science data block that is on-source ("true") or on the reference ("false"). The \texttt{bbtype} 
is also referred to as \texttt{Bbid} in the summary table of the HTP.

The \texttt{bbtypes} specified the activity of the HIFI instrument at any given time and were specific to both AOT 
and reference scheme (e.g. DBS \texttt{bbtype} have been shared between AOTs).

Level 0 contained a {\it Quality} product based on checks on the Level 0 data for telemetry issues common
to both the WBS and the HRS (in the {\it CommonTm} product). For each spectrometer, the {\it Quality} product 
also contained checks on the data frame count and quality. 

Additionally, information from the {\it HIFI Uplink} product, which contained information calculated by HSpot
concerning how the observation should have been carried out (e.g., the requested noise in the final spectra 
or the spacing between scan legs in a map) was copied to the HTP metadata. Level 0 data have units of channel 
number (''wave" scale) and counts (intensity scale).  These products are therefore of very little interest to most 
of the archive users.

\begin{table*}
\begin{center}
\caption{Listing of bbtypes which can appear in HIFI data \label{tab:bbids}}

\begin{tabular}{rrrrrr}
\hline \hline
bbtype & type & line & wbs & hrs & details \\

\hline

6004 & comb & false & true & false & WBS-Zero-Comb \\
6005 & hc & false & true & true & HIFI-Calibrate-hot-cold \\

6601 & tune & false & false & true & HRS-tune-block-aot \\
6609 & tune & false & false & false & Magnet-tuning-block-aot \\

6613 & tune & false & true & false & WBS-attenuators-block \\
6014 & comb & false & true & true & WBS-Zero-FCal \\

6021 & science & false & true & true & HIFIContOffIntegration \\
6022 & science & true & true & true & HIFIContOnIntegration \\

6031 & science & true & true & true & HIFISlowChopOnIntegration \\
6032 & science & false & true & true & HIFISlowChopOffIntegration \\

6035 & science & true & true & true & HIFILoadChopOnIntegration \\
6036 & science & false & true & true & HIFILoadChopOffIntegration \\

6038 & science & true & true & true & HIFIFSwitchOnIntegration \\
6039 & science & false & true & true & HIFIFSwitchOffIntegration \\

6042 & science & true & true & true & HIFIFastChopOnIntegration \\
6043 & science & false & true & true & HIFIFastChopOffIntegration \\

\hline

\end{tabular}
\end{center}
\end{table*}

\subsection{Level 0.5 \label{sec:levels:level05}}
At this stage, the pipeline converted the raw spectrometer output into the basic spectral elements of frequency 
and flux as a function of time. Since the two spectrometers are fundamentally different, there were two separate 
pipelines: one for the WBS spectrometer (Figure~\ref{fig:level05wbs}), and one for the HRS spectrometer
(Figure~\ref{fig:level05hrs}). Additionally, the H and V mixers are optically (semi-) independent and their output 
data in these respective chains were in their own processing thread as well.

The processing of the WBS data \citep{2003SPIE.4855..290S}  was mostly dedicated to processing the CCD 
channels of the four sub-bands. Each CCD array had pixels that were never illuminated. These pixels determined 
the "dark" for the array. As part of the standard operations, zeros and comb measures were obtained together. 
A zero measurement was constructed by interpolation in time between two measures where possible, and 
subtracted from the fluxes. Generated from the 10 MHz master oscillator (Local Oscillator Source Unit, LSU), the 
comb signal is a series of stable frequencies at 100 MHz steps. This comb signal is used to assign a frequency 
scale to the WBS CCD channels. A quality measure of this processing was retained in the {\it Quality} product for 
the WBS data. Other quality information relative to the CCDs was also retained.   

The Level 0.5 pipeline for the HRS was the implementation of the signal processing described in 
\citep{2004NewA....9...43B}.  { The HRS had 4080 autocorrelation channels that were able to provide up to four sub-bands per polarisation. 
The observer had a choice of configurations of the HRS depending on the desired spectral resolution:  wide band, low, normal and high resolution modes.  It was possible 
to place each sub-band independently anywhere within the 2.4 or 4 GHz IF.  
The HRS processing split the telemetry into sub-bands depending on the HRS mode used.   The autocorrelation functions per sub-band were  fast-Fourier transformed resulting
in flux at each channel and sub-band.  Finally, frequency columns for each sub-band were created.}

The result of the Level 0.5 pipeline was a time series of integrations per back-end spectrometer in instrumental 
counts as functions of frequency. 

These products were removed from the observation as soon as the Level 1 products were successfully generated. 
They can, however, be regenerated interactively if needed. 

The only product retained in the Level 0.5 was the quality information of the WBS-H and WBS-V comb, and zero 
quality checks which are stored in a {\it Quality} product.

\subsection{Level 1.0 \label{sec:levels:level10}}
The purpose of the Level 1 pipeline was to flux-calibrate the HIFI data using the internal load measurements, and 
to apply the referencing scheme that was used for the observation. Additionally, the observed frequencies were 
placed into the LSR velocity frame for fixed targets, and in the reference frame of the moving targets for solar system 
objects (SSOs). At this level of processing, integrations taken at different times were then combined. The 
overall flow of the Level 1 pipeline is shown in Figure \ref{fig:level1} and is discussed below.

The quality product at Level 1 contained the results of the phase 
checks made by the Level 1 pipeline for each spectrometer that was used in the observation.

The processing steps taken during the Level 1 pipeline were as follows:
\begin{itemize}
	\item \texttt{checkDataStucture, checkFreqGrid, checkPhases}: the first part of Level 1 pipeline performed
	a series of sanity checks on the data structure, frequency grid, and phases. The reported chopper positions 
	were checked to lie within the appropriate ranges. This step was used to verify that the data structure 
	belongs to the specified AOT and raised a quality flag reporting discrepancies. \texttt{checkFreqGrid} 
	created a product called {\it FrequencyGroups} in the {\it Calibration} product which contained the analysis 
	of the LO groups. \texttt{checkPhases} performed an analysis of the patterns found in the sequence of science 
	and load measurements to be consistent with the observing mode, specifically, whether the correct sequence 
	of chopper positions, buffer values, ON/OFF positions, or LO frequencies could be recovered in the observed 
	data. Quality flags were raised when issues were identified.
	\item \texttt{doFilterLoads}: this step provided a means to reduce the standing waves that could be 
	introduced from the internal loads.  { This step could reduce the 90-100 MHz standing waves in bands 1 and 2 on 
	strong continuum sources, but 
	had limited success (even degradation) in other bands.} In the default pipeline, this task was disabled. 
	\item \texttt{mkFluxHotCold}: this calculated the calibration to the antenna temperature scale $T_\textrm{A}^\prime$ 
	(that is, the antenna temperature before correction from the rearward beam efficiency losses) 
	\citep{1981ApJ...250..341K} based on the hot/cold load measurements and created a load-calibration object. 
	The calibration held two objects, one was the bandpass that calibrated to the \Tprime ~temperature scale:
	\begin{equation}
		\textrm{bandpass}= (L_h-L_c)/[(\eta_h + \eta_c - 1) \times (J_h - J_c)]
	\end{equation}
	the second object was the system noise temperature based on the load integrations and the measured load 
	temperatures ($T_h, T_c$) ,
	\begin{equation}
	T_{\textrm{sys}} =  {{[(\eta_h+\textrm{Y} \times \eta_c-\textrm{Y}) \times J_h-(\eta_h+\textrm{Y} \times \eta_c-1) \times J_c]} \over 
		{\textrm{Y}-1}}
	\end{equation}
	where $\eta_h ~\textrm{and}~ \eta_c$ are the coupling efficiencies towards the hot and cold loads, $J_h ~
	\textrm{and} ~	J_c$ are the Rayleigh-Jeans equivalent radiation temperatures of the hot and cold loads 
	at ${T}_{h} ~ \textrm{and} ~ {T}_{c}$, respectively, and $ L_h ~\textrm{and} ~ L_c $ are the load integrations 
	in counts. The factor Y is the ratio of the load integrations, $\textrm{Y} = L_h/ L_c $. The variables, $L_h, 
	L_c, J_h, J_c, \textrm{Y}, {T}_\textrm{sys}$, and the bandpass are all functions of frequency and hence are arrays.
	\item \texttt{doChannelWeights}: this calculated and applied radiometric weighting to each channel.  
	$ w = t_{integr} / {{T}_\textrm{sys}^2}$ , where $t_{integr}$ is the duration of each integration. For the 
	purpose of the data weights,  	
	${{T}_\textrm{sys}} $ needed to be interpolated in time between any two load integrations or the nearest in time 
	if no bracketing data existed. The weights were further smoothed over 20 channels.
	\item \texttt{doRefSubtract}: the purpose of this step was to perform the first difference of the double-difference 
	scheme, by subtracting the reference signal. This step ass highly dependent on the referencing scheme.
	\begin{itemize}
		\item For the DBS scheme, the chopped reference was subtracted by assuming an ABBA pattern (here,
		A and B means REF and ON respectively).
		\item For the fastDBS scheme, the chopped reference was subtracted by assuming an ABAB pattern.
		\item For the FSW scheme, the integrations that were shifted in frequency by the throw 
		were subtracted. This scheme assumed an ABBA pattern.
		\item For the Load Chop scheme, the integrations on the cold load were subtracted from the target 
		integrations. This scheme assumed an ABBA pattern.
		\item For PSW schemes, this step was not performed. No fast reference was taken in these schemes.
	\end{itemize}
	 \item \texttt{mkOffSmooth, doOffSubtract}: when the observation was taken at a dedicated OFF position, 
	 the OFF signal was subtracted from the source signal. DBS observations always had OFF source positions 
	 3\arcmin ~on either side of the source given by the telescope slew. In the case of Load Chop or FSW 
	 referencing schemes, the OFF data were smoothed by a width parameter depending on the HIFI 
	 band used. Table \ref{tbl:smoothwidths} lists the smoothing widths per LO of the OFFs in this case.  { Based on 
	 calibration tests, the smoothing
	 widths were chosen to minimize the noise added back into the spectrum via the \texttt{doOffSubtract} step.} 
	When the observation
	 was an OTF map or a PointSpectrum mode with PSW referencing, the OFF samples were averaged and
	 interpolated to the time of the ON observation, and subtracted. A calibration product called {\it Baseline} 
	 was created in the {\it Calibration} product and contained the baseline from the OFF positions.
	 \item \texttt{doFluxHotCold}: this applied the flux calibration based on the load measurements by dividing the 
	 signal by the bandpass interpolated to the time of the signal integration. The bandpasses were determined
	 for each hot/cold measurement as described for the \texttt{mkFluxHotCold} step.  These hot/cold 
	 measurements typically occurred at the beginning and the end of an AOR, but they could also occur during an AOR
	 (depending on the stability of the mixer). The resulting bandpasses were linearly interpolated to the time 
	 of the source integration, or if the source integration was not bracketed by a hot/cold sequence, the 
	 hot/cold sequence nearest in time was used. We  note that for the observing modes Load Chop and Frequency Switch,
	 \texttt{doFluxHotCold} was applied before \texttt{mkOffSubtract} and \texttt{doOffSubtract} in order to reduce 
	 the noise.
	 \item \texttt{doVelocityCorrection}: taking into account the satellite motion with respect to either the local 
	 standard of rest (LSR)  or an SSO, this task corrected the frequency scale to either LSR or the 
	 appropriate SSO velocity.  This task used the full relativistic corrections.
	 \item \texttt{doHebCorrection}: when the observation was made  in band 6 or 7, the task applied the correction for the 
	 electrical standing waves \citep{2014AIPC.1636...62K}. The corrections were contained within the 
	 \texttt{Downlink} section of the {\it Calibration} product.
	 \item \texttt{mkFlagSummary}: this produced a summary table of any flags (per channel or per integration) 
	 that were raised during the Level 1 processing. This information was fed into the Leve1 {\it Quality} 
	 product as {\it FlagSummary}.
\end{itemize}

At the end of the Level 1.0 pipeline, all science spectra were flux calibrated to the antenna temperature
$T_\textrm{A}^{\prime}$ and the intermediate frequency was adjusted to V$_{LSR}$ reference (still in 
MHz). At this stage, separate integrations were not co-added, and calibration spectra (e.g. hot / cold 
loads) were still present in the HTP.

\begin{table}
\begin{center}
\caption{Off spectra smoothing widths for both H and V polarisations \label{tbl:smoothwidths}}
\begin{tabular}{llll}
\hline \hline
band&frequency&LoadChop&FSwitch \\
&GHz&MHz&MHz \\

\hline
1a-5b & 480 - 1280  & 9.0 & 11.0 \\
6a & 1420 -1457 & 30.0 & 11.0\\
6a & 1457 - 1459 &10.0 & 11.0\\
6a & 1459 - 1522  & 30.0 & 11.0\\
6a & 1522 - 1570 & 18.0 & 11.0\\
6b & 1570 - 1655 & 18.0 & 11.0\\
6b & 1655 - 1710 & 30.0 & 11.0\\
7a-7b & 1710 - 1910& 18.0 & 11.0\\
\hline
\end{tabular}
\end{center}
\end{table}

\subsection{Level 2.0 \label{sec:levels:level20}}
The Level 2 pipeline (Figure~\ref{fig:level2}) contained spectra for each spectrometer that was used in the observation. 
Level 2 data were converted into \Tstar ~scale (in K) and into sky frequency (GHz). Owing to the DSB frequency 
degeneracy, products were generated on both the USB and LSB frequency scales. Spectra were averaged together for each  spectrometer for each LO setting, and for each spatial position in the observation. This resulted in a single 
spectrum (for each spectrometer) for point-mode observation, in an individual LO setting for spectral scans, and 
individual spectra per position and LO setting for maps.

The processing steps taken during the Level 2 pipeline were as follows:
\begin{itemize}
        \item \texttt{mkRef}: this step extracted emission in reference positions. For DBS modes, this step 
        calculated the difference in emission between the two chop positions. The resulting spectra gave 
        a rough idea of whether the proposer chopped onto a region with emission. This step populated 
        the calibration \texttt{pipeline-out} with the OFF spectra.
	\item \texttt{doCleanUp}: Up to this point, very little had been removed from the time-line 	
	products. This initial step to the Level 2.0 pipeline removed spectra used for calibrations 	
	(e.g., WBS comb spectra). 
	\item \texttt{doAnntennaTemp}: this converted data into antenna temperature \Tstar ~by dividing all spectra 
	by the forward efficiency { $\eta_l$.  We note that $\eta_l$ is the combination of radiation and rearward spillover and scattering losses as defined by \citet{1981ApJ...250..341K}}. For all HIFI bands, this efficiency is 0.96.
	\item \texttt{mkSidebandGain}: this created a calibration object that contained the band 
	and IF specfic sideband gains that were  to be applied to the given observation \citep{sidebandratio}.	
	\item \texttt{doSidebandGain}: this calculated upper and lower sideband spectra by dividing by 
	the sideband gain determined in the \texttt{mkSidebandGain} step. This step created
	spectra that are listed as WBS\_USB,  WBS\_LSB, HRS\_USB, HRS\_LSB, but note that 
	the USB and LSB determinations are data taken from the intermediate frequency, and a LSB 
	spectrum (for example) may have a USB spectral feature in it. Truly separating sidebands 
	requires multiple LOs to have been used, and the deconvolution task to have been run
	(see \HIFIHB section 5.8.4). Furthermore, we note that sideband-gain correction is only applicable to 
	line emission from a given sideband, so that the continua of spectra after this correction were typically 
	a factor 2 higher than the true single-sideband continuum (see \HIFIHB).
	\item \texttt{convertFrequency}: this step converted the frequency scales from IF (in MHz) into 
	USB or LSB (in GHz) scales. For HIFI bands 1-5:
	\begin{equation}
		\nu_{USB} = \nu_{LO} + \nu_{IF} ,~ \nu_{LSB} = \nu_{LO} - \nu_{IF}
	\end{equation}
	For HIFI bands 6 and 7:
	\begin{equation}
		\nu_{USB} = \nu_{LO} + \nu_{up} - \nu_{IF} ,~ \nu_{LSB} = \nu_{LO} - \nu_{up} + \nu_{IF}
	\end{equation}	
	where $\nu_{up}$ is an additional up-conversion frequency given by 10.4047 GHz for horizontal 
	polarisation and 10.4032 GHz for vertical polarisation. 
	\item \texttt{mkFreqGrid}: this step created a frequency grid with uniform spacing. For WBS this was
	set at 0.5 MHz. The frequency grid spacing for HRS is found in the metadatum \texttt{channelSpacing}. 
	\item \texttt{doFreqGrid}: This step resampled the original fluxes and weights onto the uniform 
	frequency grid provided in the previous task. The task used the Euler scheme and conserved total 
	power. As noted above, the data were separated into upper- and lower-sideband products. The 
	result of this frequency re-sampling step may be that the upper and lower sideband spectra now differ 
	in array length.  
	\item \texttt{doAvg}: when the observing mode was not a spectral map, all spectra at a given LO and sideband 
	were co-added. This step created products per spectrometer, polarisation and sideband. The averaging of 
	spectra was performed without checking for standing waves or baseline issues. In some cases, it was not 
	necessarily beneficial to co-add all the spectra. For this reason, an "expert-generated" User-Provided Data 
	Product of non-averaged Level 2 spectra is provided through the 
	HSA\footnote{\href{http://www.cosmos.esa.int/web/herschel/user-provided-data-products}{http://www.cosmos.esa.int/web/herschel/user-provided-data-products}\label{updp_fn}}.
	\item \texttt{doHpbw}: when the observing mode was a spectral map, then the half-power beam 	
	width was added to the metadata of the observation and was used to define the convolution 
	kernel during the re-gridding process in the Level 2.5. This guaranteed that the spectral map was 
	made with the same assumptions that were available to the observer at the time of planning the 
	observation.
	\item \texttt{checkPlatforming}:a quality check was performed on all WBS spectra to determine whether
	adjacent sub-bands were consistent with each other. Each WBS sub-band overlapped with its 
	neighbour. When the difference in flux values in the overlap region was more than 1.3 times the 
	noise, a \rm{\texttt{platforming}} flag was raised and reported in the product as well as 
	in the quality summary. We note that this check is all or nothing, that is, if only one spectrum 
	showed the platforming, the flag was raised on the entire observation.
	\item \texttt{mkRms}: the pipeline also calculated what the noise was in the final spectra and 
	compares this with the predicted HSpot results. The resulting value was placed in the 	
	{\it Trend} product of the observation. This step produced quality flags when the noise of polarisation 
	data differed too much (greater than 20\%) or when the combine noise was significantly higher (>30\%) 
	than the original HSpot prediction. Either of these flags appeared in the {\it Quality Summary} 
	report for the observation only when the limit was exceeded. The noise was determined, ideally,
	on a "line-free" portion of a baseline-corrected spectrum. This automated step had to identify and 
	mask any spectral feature that appeared as a spectral line and removed any baseline drifts. 
	Furthermore, the spectra were smoothed to the desired smoothing width provided by the 
	observer to plan the observation.
	\item \texttt{doChannelFlags}: using data from the {\it Calibration} tree, this task flagged data in which spurs 
	were known or suspected to exist. This step was necessary for higher quality spectral survey 
	results as these flags forced the deconvolution to ignore flagged data. The flags were also added 
	to point spectra and spectral maps, but this was without consequences for any further data processing.   
	\item \texttt{mkUncertaintyTable}: this step generated products containing a table of random and 
	systematic uncertainty values for the observation based on uncertainties for sideband ratio, optical 
	standing waves, hot/cold load coupling, hot/cold load temperature, planetary model, aperture, 
	and { main beam efficiency}. The uncertainties are listed as percentages. A table is available for each spectrometer, 
	polarisation, and sideband. The uncertainties are calculated based on \citet{ossenkopf2015}.
\end{itemize}

The results of the Level 2 pipeline for non-mapping mode data were coadded spectra. These spectra 
are listed either on the USB or LSB frequency scale. The flux calibration is in \Tstar . 

\subsection{Level 2.5 \label{sec:levels:level25}}
The Level 2.5 pipeline processed the Level 2 data products into final nearest-to-science-ready products 
as possible, in operations that depended on the observing mode. The products created at the end of the 
Level 2.5 were then used to create the stand-alone browse products (Section \ref{sec:browse}).  

In the historical evolution of the pipeline, Level 2.5 processing was defined somewhat flexibly for each 
\Herschel instrument to carry out value-added enhancements to the essential deliverables generated by the Level 
2.0 pipeline. In other words, Level 2.0 products were absolutely required and scoped as such in terms 
of development resources starting during Instrument Level Testing (ILT) in the pre-launch phase, whereas 
further processing needed to reach science-ready products that could be achieved in other known 
processing environments available to astronomers was given lower priority. However, with the maturing 
of the pipeline and accumulation of products from all of HIFI's available Observing Modes in the HSA as 
the mission progressed from Performance Verification into the Routine Phase, it became clear that Level 
2.5 processing within the same HCSS environment was more and more necessary. The Level 2.5 processing
step was created in order to validate both the lower level pipeline and the performance of the Observing 
Modes, and to provide end users with the products which turned out to be much more demanding than initially 
estimated. 

Export of all Level 2.0 products along with necessary calibration tables for certain processing (such as in 
the construction of spectral cubes) had many practical limitations, and small program teams also had limited 
resources to develop or adapt other software packages, or manage the high volume of data being produced 
by each observation up to Level 2.0. This was generally true for users of any of the three \Herschel instruments; for HIFI 
the resource demands and complexities were greatest for spectral scans and spectral maps. As a consequence, 
Level 2.5 steps were introduced relatively early in the mission, first for the spectral maps and spectral scans 
with robust algorithms for producing spectral cubes and sideband-deconvolved 1D spectra, and later for Point 
Mode spectra involving more simple operations requested by users. While the added value of this processing 
was clear to HIFI scientists for iterative improvements to the underlying pipeline and AOT logic, the effort was 
further encouraged by the Herschel User Group and the Data Processing User Group.     

Since the tasks that created Level 2.5 products were considerably more algorithmically complex than 
the individual tasks in the lower pipeline, especially for Mapping Mode and Spectral Scan observations, it 
is worthwhile to elaborate on their implementations.  

\subsubsection{Point Mode} 

Level 2.5 processing of Point Mode spectra was essentially a conversion of formats for ease of handling by 
end users. The spectra were sub-band stitched (\texttt{doStitch}), folded (in the case of Frequency Switch) into 
a single 1D spectrum, and converted into the so-called \texttt{SimpleSpectrum} format, again on a spectrometer,  
sideband, and polarisation basis. WBS spectra had IF sub-bands with overlaps of roughly 150 MHz and were
always stitched. Each sub-band was cut at a cross-over point that was generally the mid-point of the overlap; other 
options were available in HIPE. This resulted in a single spectrum, which simplified its handling by end users, but 
it lost information on potential sub-band-specific effects such as platforming (see {\it HB}). HRS data were stitched 
only when the sub-bands substantially overlapped  in frequency, which was not always the case since this depended on 
the requested resolution mode and placement of the HRS sub-bands by the user during observation planning. 

The stitched spectral data were populated into a so-called \texttt{Spectrum1d}, with columns for wave (frequency), 
flux (antenna temperature), flagging, weights, and segment number (set to unity after stitching). The data were
contained in the HIFI data tree as \texttt{SimpleSpectrum} products, which are basically metadata wrappers around 
the \texttt{Spectrum1d} with links or ``bindings", and pointers to other parts of the data tree. The \texttt{SimpleSpectrum}
product definition was intended to allow flexibility in HIPE software applications for spectra produced by all 
three instruments.   

\subsubsection{Mapping Mode} 

Spectral mapping observations carried out at various single-dish ground-based receivers are typically provided 
to users as spectral cubes built up from planes of signal projected into celestial coordinates at each frequency 
channel, thus there do exist different software environments that might conceivably carry out the construction 
of the cubes with input Level 2.0 HTPs. The conversion of formats from the very instrument- and HCSS-specific 
HTPs into required file types such as standard FITS or CLASSFITS (for example) is non-trivial, however. 
Furthermore, in order to facilitate all of pipeline and AOT logic optimisations iteratively and self-consistently 
within the same processing environment, and fill a requirement by end-users for this key science product, the 
construction of spectral cubes was the first extension of the Level 2.0 pipeline into Level 2.5, and initially 
integrated into the SPG already during the Performance Verification phase (although still defined as Level 
2.0 at that time).  

The Level 2.5 processing for mapping observations began with sub-band-stitching of the Level 2.0 HTPs, 
that is, the individual spectra corresponding to different map points on the source were treated as Point Mode 
spectra and stitched as described above. Similar to Point Mode observations, stitching was always made 
to WBS spectra, and made to HRS spectra (if present) only when the sub-bands overlapped. The Level 2.5 HTPs 
are provided as products in the HIFI data tree, and constitute the inputs for the pipeline \texttt{doGridding} 
task that creates spectral cubes on a spectrometer (WBS/HRS), sideband (USB/LSB), and polarisation 
(H/V) basis. 

The basic purpose of the \texttt{doGridding} task was to place all mapped on-source data from a Level 2.5 
HTP into a regular spatial grid as a function of frequency. The  algorithm is very similar to that used for 
mapping observations at the 30 m IRAM telescope, in which the Globalized Sinusoid (GLS) projection 
method \citep{2002A&A...395.1077C} provides the convolution of individual spectral readouts into effective 
``pixels'' (sometimes referred to as ``spaxels'' for spatially synthesised pixels in non-imaging devices). In 
the case of HIFI, the pixel sizes were usually determined by stored parameters that reflect how the readout 
pattern was commanded  on the sky, specifically, the number of scan lines, number of readout points per 
scan line, and the requested sampling density (Nyquist, half-beam, 10$''$, 20$''$, or 40$''$ readout spacing). 
This information was not available for all observations, mainly from the early part of the mission, and in these 
cases, the task attempted  to determine the nominal sampling from the attitude assignments to the spectra in 
the Level 2.5 HTPs. In both cases, after the initial-guess dimensions were determined, all spectra which fall 
spatially within the convolution kernel were convolved to fill each rectangular pixel.    

The default convolution was an azimuthally symmetric Gaussian kernel based on the half-power beam width 
(HPBW). Values for the HPBW are provided in a look-up table in the observation's Calibration Product at 
several frequencies in each mixer band (summarized in Table~\ref{tbl:spotFreqs}) based on flight beam 
calibrations \citep{HIFI-ICC-RP-2014-001}, and interpolated to the LO frequency of the observation. The 
2D Gaussian function in $x$ and $y$ positions (i.e., indexed map position) is expressed as

\begin{equation}
\label{eq:gaussfunc}
G(x,y) \propto \exp^{-(({x}^2+{y}^2)/2{{\sigma}^2})}
\end{equation}

The pipeline used a kernel sigma $\sigma_k  = 0.3 \times \sigma_{\rm B}$ where 
$ \sigma_{\rm B}$  =  HPBW / (2 $\sqrt{(2 ln2)}$). { The factor 0.3 results in a slightly larger (4\%) final beam in the regridded data cube.}  While the spacings $\Delta x$ and $\Delta y$ between 
map points were set to be equal in the AOT logic, determined only by the requested sampling and the beam 
size at the associated LO frequency, deviations from the planned path of the telescope occurred because neither 
the actual nor reconstructed telescope points were errorless.  

The contribution of data points near a spatial pixel was determined by the Gaussian-weighted average of all
data within an influence length. Data points that were within an influence length of $3 \times \sigma_k$ in 
both $x$ and $y$ from the centre of the pixel were convolved. The weight factor $G_n (x,y)$ for the $x, y$ 
pixel from data point ${n}$ at the grid position $x_n,y_n$ is:

\begin{equation}
\label{eq:filterparam}
G_n (x,y) = \exp^{-1/2 (x-x_n)^2 / \sigma_k^2}~ \times~ \exp^{-1/2 (y-y_n)^2 / \sigma_k^2}
\end{equation}

\noindent where $x_n$ and $y_n$ are the positions of the observed data point $n$.  

The convolution combines data within an influence length. When $N$ data points fall within this length for a pixel 
at ($x,y$), the flux is given as convolution of the data fluxes, $F_n$:
\begin{equation}
F(x,y) = \Sigma_{n=1}^N (G_n(x,y) \times F_n) / \Sigma_{n=1}^N G_n(x,y)
\end{equation}

\begin{figure}[h]
\includegraphics[width=0.4\textwidth]{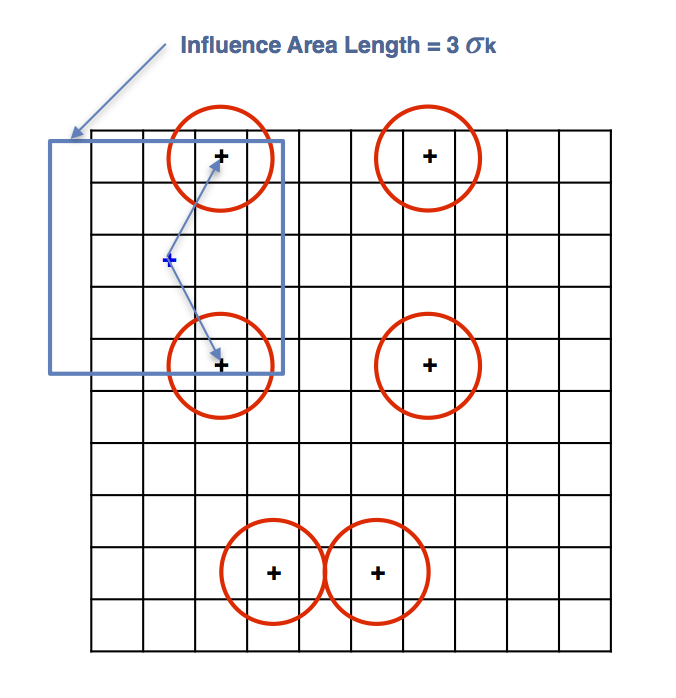}
\caption{Schematic representation of the convolution of spectral map points onto a regular grid by the
\texttt{doGridding} task. The pluses represent map readout points with the HPBW represented 
by red circles, and the blue squares represent an influence area.}
\label{fig:gaussianfilterfig}
\end{figure}

Figure \ref{fig:gaussianfilterfig} shows an example of six beam positions (red circles) on a grid, where two 
positions are within the influence area (blue square).  

\begin{figure}[h]
\includegraphics[width=0.4\textwidth]{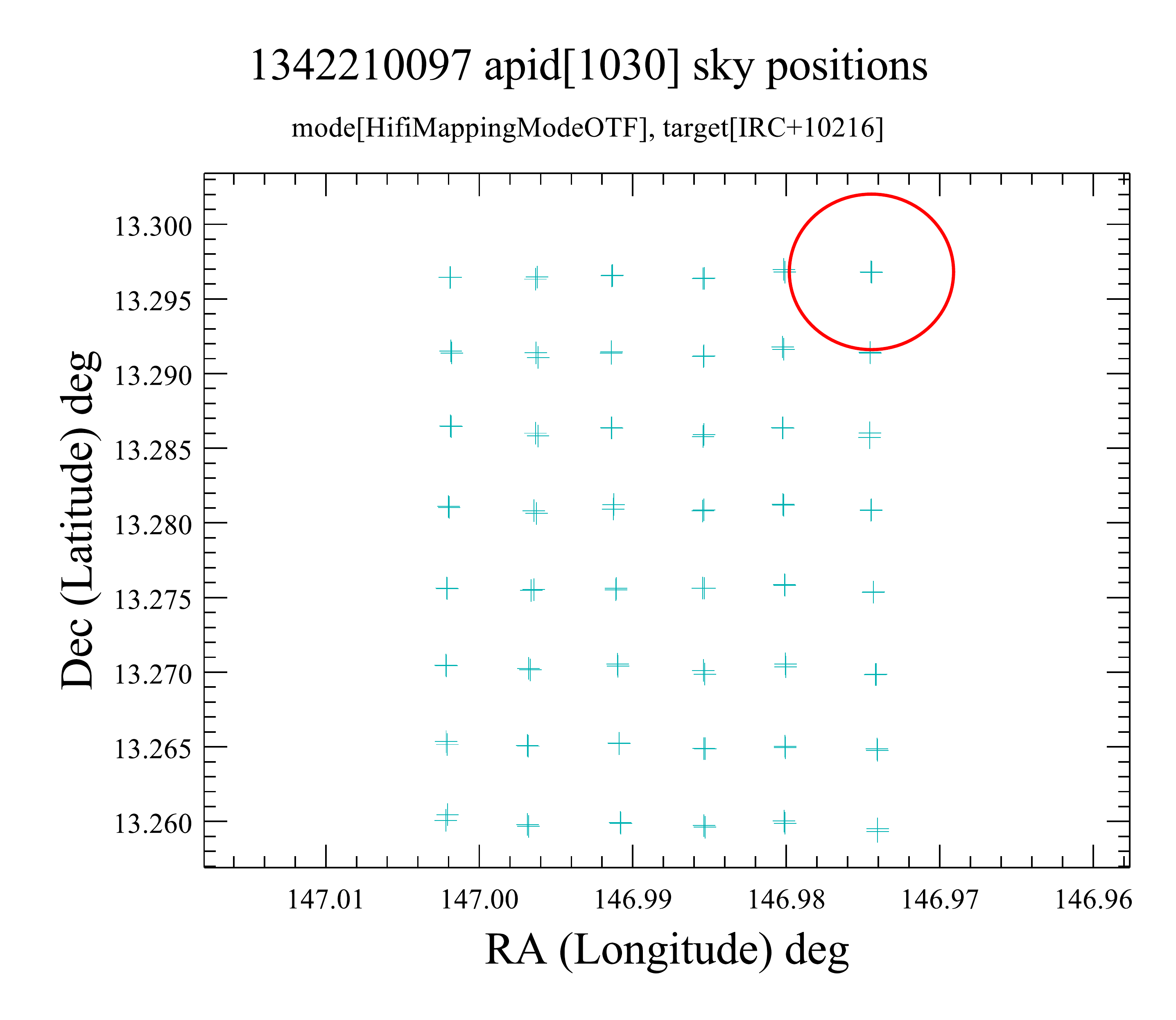}\\
\includegraphics[width=0.4\textwidth]{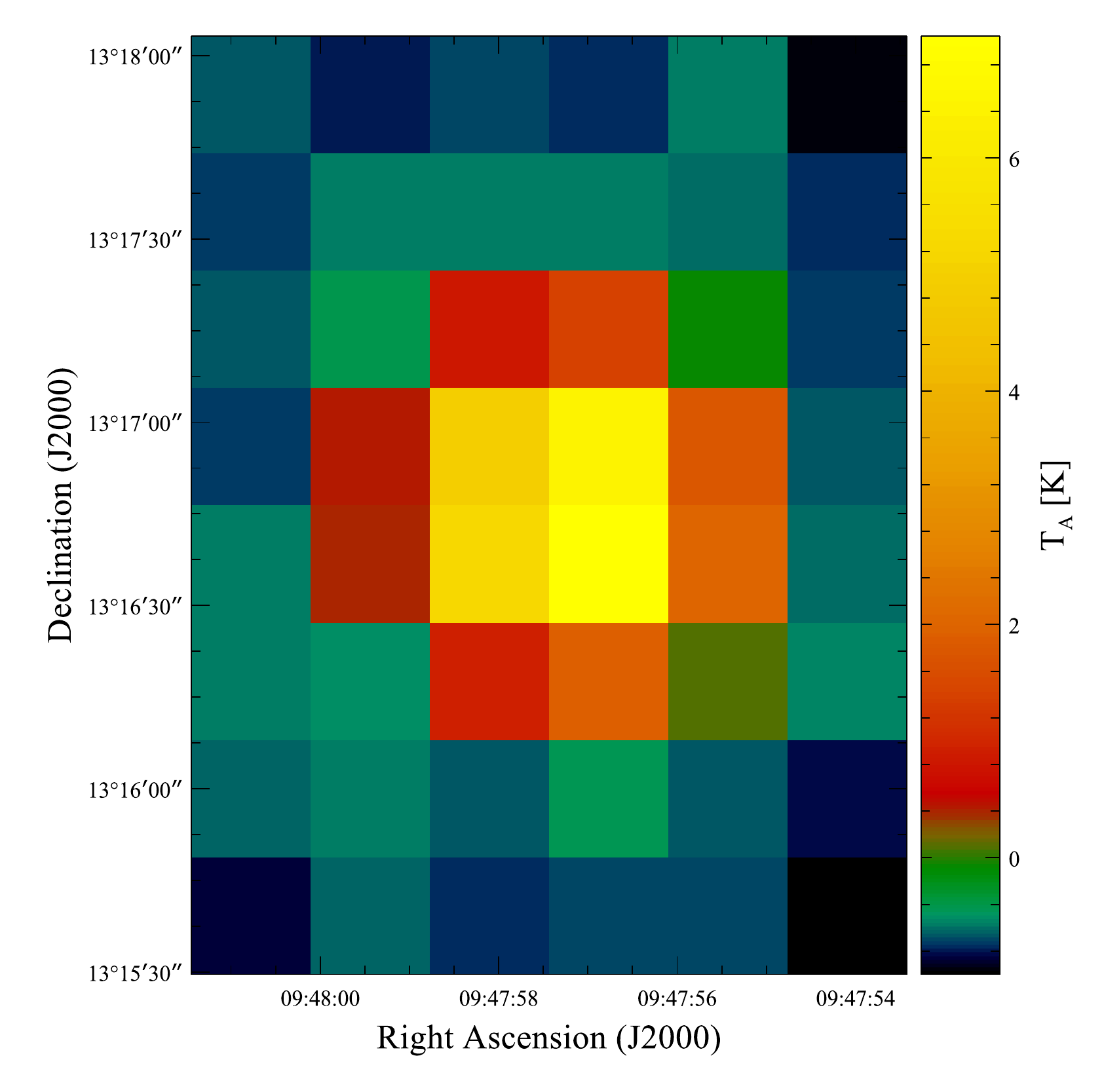}\\
\caption{{\it{Top}}: Sky positions for map points of an OTF map (obsid 1342210097) of IRC~+10216 with
the LO tuned to USB coverage of CO $J=5-4$ in band 1b. The pluses represent the attitudes read 
from the PointingProduct and assigned to each spectrum data-set in the Level 1 pipeline. Two map cycles 
using half-beam sampling were carried out in this observation. The size of the beam (HPBW = 37$''$.5) 
is shown in red. {\it{Bottom}}: The map gridded into square pixels for the same observation. Each pixel is 
19$''$.5 $\times$ 19$''$.2, i.e., not perfectly square because of overall pointing errors that do not average 
out in the entire map.}    
\label{fig:mapfig}
\end{figure}

Each spectrum from the input HTP was not necessarily unique in terms of its celestial coordinates; more 
than one spectrum per map point was produced in cases where the observation was carried out over 
more than one mapping cycle, driven by baseline noise goals during observation planning. No two 
spectra taken at the same map point had identical coordinates because of uncertainties in the actual and 
reconstructed telescope pointing, but the differences for spectra taken at the same point were usually 
small compared to the HPBW and tended to be noticeable mainly in maps taken with the OTF mode. An 
example is shown in Figure~\ref{fig:mapfig}. The positional differences between spectra at the same 
map point, as well as slew deviations from the ideal path of the telescope during mapping such as the 
OTF ``zig-zag'' or ``hula dance'' effect (Morris 2011, 2015), were taken into account in the weighting filter 
profile during convolution. These deviations have had some important consequences:

\begin{itemize}
\item Throughout most of the mission, the pointing reconstruction did not accurately reflect the anomalous 
telescope performance of the line-scan-with-off pointing mode used for OTF mapping, which led to the 
apparent ``zig-zag'' or alternating shift of pixels on each scan line in the gridded spectral cubes. The 
effect was strongest in HIFI's high-frequency bands (smallest beams). The effects were greatly reduced 
after the drift from the four gyroscopes on the telescope was characterised more accurately, leading to 
a more accurate pointing reconstruction and thus signal weighting during cube construction (in most cases). 
\item The gridded pixels were not always of equal size on each axis because of the general pointing performance 
uncertainties. In many observations the effects were essentially averaged out, and pixels were very close to
square, the differences being much smaller than a fraction of the beam size. Some small maps that had 
only a few map points, or maps that were taken in the early part of the mission, were particularly susceptible to variations in angular angular length in one or the other dimension. The example observation shown in the bottom panel of 
Figure~\ref{fig:mapfig} has slightly rectangular (non-square) pixels for this reason, where the slight deviations 
seen on close inspection in the upper panel are not completely averaged out in both dimensions on the sky. 
\end{itemize}

We note that the HIFI beams were only used for weighting in the convolution kernel, and for this purpose 
the approximation of the beams as 1D symmetric Gaussian profiles was adequate. The \texttt{doGridding} 
task does {\it{not}} have any capability to deconvolve the beam shapes from the signal in the constructed 
cubes. This operation was considered a worthwhile augmentation of the task in interactive applications, but 
ultimately this functionality was not developed because of system software stability risks and resource limitations.    

At this stage, the processed spectral cubes were added to the HIFI data tree under a Level 2.5 ``cubesContext''.
The cubes were readily operated on by tasks in HIPE and exported to FITS. A second set of cubes was produced 
in the SPG with the \texttt{doGridding} task for maps that were taken at a non-zero rotation angle (i.e., 
not square with R.A. and Dec). In this case, the GLS algorithm was applied at the specified rotation angle to 
project the map into a filled square map with generally non-square pixels correctly projected in celestial 
coordinates such that signal was conserved (theoretically) compared to the pixels constructed at native 
orientation. These cubes, similarly generated on a spectrometer, sideband, and polarisation basis, were 
populated into a ``cubesContextRotated'' Java construct.  

It should be noted that the \texttt{doGridding} task in the SPG assumed equal quality of all spectra within the 
input HTP. However, when any frequency channel(s) of the input spectra was flagged for quality problems, such 
as possible spurs, saturation, or unruly conditions of the WBS CCD pixels spreading across many channels, 
the task gave zero weighting to the affected channels during signal convolution. This could  result in an 
\texttt{NaN} assignment to the signal at the associated frequencies in some or all pixels, unless more than 
one map cycle was taken and the problem was transient and not systematic over the entire observation.   

The \texttt{doGridding} Application Program Interface (API) included extensive top-level parametrisation to 
allow flexible user interaction with the task in HIPE. This was somewhat different in the lower level pipeline 
tasks --- although most of the tasks were parametrised for customisable processing in HIPE, intervention in 
the Level 0 to 2.0 pipeline by users has been rarely needed (and in these rare cases usually involved 
rerunning the pipeline in a default configuration that used updated calibrations or bug fixes between major 
releases of the HCSS or bulk reprocessings in the HSA). The most common customisations of spectral 
cubes involved resampling to a different pixel size, to a finer grid when signal-to-noise ratios around a target 
spectral line were high, for example, or using different filter parameters to mimic the beam properties in 
mapping data obtained with other instruments, or matching the WCS of another observation to a HIFI 
mapping observation with spatial overlap. The task also had to be re-run in order to reach optimum baseline 
noise performances using input HTPs constructed with the \texttt{mergeHtps} task to combine data sets 
from the two polarisations, or from different observations. These customisable features were exploited 
to produce the mapping HPDPs available from the HSA (see footnote \ref{hpdp_fn}).

\subsubsection{Spectral Scan Mode} 
The concept of the Spectral Scan observing mode was that sufficient redundancy was built in sky frequency 
coverage in order to resolve the double-sideband degeneracy intrinsic to the HIFI detectors. The higher 
the redundancy, the more accurate the single-sideband spectrum that was reconstructed from the double 
sideband Level 2 spectra. The sideband deconvolution was performed for each polarisation, following the 
conjugate gradient method \citep{2002A&A...395..357C}. In this process, all artefacts present in the 
Level 2 spectra inputs were propagated into the final deconvolved spectrum. This was particularly critical for 
narrow spectral spurious features, which polluted the end results. This required careful 
masking of those spurious artefacts as one of the pipeline steps before the sideband deconvolution. Such 
masks were manually assigned for each individual spectrum involved in one of the nearly 500 Spectral 
Scans taken with HIFI. The masks were stored as calibration files and attached as flags to the spectra, and 
they were subsequently used to identify spectral channels to be ignored in the construction of the deconvolved 
spectrum.

Another fundamental parameter of the sideband deconvolution was the sideband gain to be applied to the respective 
sidebands. In the HIFI pipeline, this gain was tabulated as a calibration parameter and was applied to the Level 2 
spectra as a function of their LO frequencies. This means that the gain had to be undone by the deconvolution 
algorithm, and then applied to the different spectral building blocks used to construct the single-sideband solution. 
In an earlier version of the deconvolution task coded for HIFI, an option was implemented that would attempt to 
actually fit the sideband gain based on the redundant information held by the Level 2 spectra. This approach, 
however, proved to provide inaccurate gains and was removed in the end. Instead, the chosen approach was to 
use sideband gains determined independently \citep{sidebandratio} as preset value to then apply in the task.

It should be noted that the deconvolution code implemented for HIFI was initially inspired by the code implemented in 
the GILDAS/CLASS software. Features such as channel masking and ad hoc sideband gain usage were, however, 
added as extra functionalities to the initial CLASS implementation. It is now possible to benefit from
the masking effort in HIFI spectra imported into CLASS, for instance, see \cite{bardeau2015a} for further details.

The above processing guaranteed the best possible deconvolved solution from the calibrated Level 2 spectra, 
which was provided as Level 2.5 spectra. However, this does not cover any residual instrument artefact that could 
manifest as baseline distortion or ripples. For this purpose, a fraction of the HIFI Spectral Scans was further 
processed using dedicated baseline cleaning. These spectra were provided as Highly Processed Data Products 
(see footnote \ref{hpdp_fn}).

\subsection{Testing and validation of the pipeline}
\label{sec:concepts:test}
Major components of the pipeline (Level 0 and Level 0.5) have been in active use since the pre-launch instrument 
models of HIFI; hence, all the products have a metadatum named \texttt{modelName} with the value \texttt{FLIGHT}. 
Before launch, Level 0 and Level 0.5 had been tested and validated, and extensive unit tests for all Level 1 and 
Level 2 processing tasks had been available and were based on regression data constructed in accordance 
with the expected behaviour of the instrument. However, only with the full integration on the  satellite, could the 
later stages (Level 1 and Level 2) of the pipeline be truly tested. These stages were validated as part of the 
Performance and Verification phase of the HIFI mission and led to the release of standard modes of observing 
with HIFI\citep{AOTReleaseNote2011}. The performance of the HIFI instrument is described in \citet{teyssier2015}. 
Further testing and validation occurred periodically as the HCSS software was upgraded. The pipeline results also 
underwent significant testing during the mission for every major software upgrade of the \Herschel ground segment. 
This verification comprised testing of new functionalities as well as regression testing. All modes were tested. Only 
when all problems had been fully addressed would the new version of the software be accepted to be released.

\section{Assessing data quality}
\label{sec:quality}
There are two main questions regarding data quality: (1) whether each processing step { performed} within nominal 
bounds and, (2) whether the AOR fully mitigated the instrument instabilities. The first question was monitored during 
the processing by retaining quality measurements during the processing of each step. Section \ref{sec:qualityreport} 
describes quality flags when one of the processing steps was out of  limits. 

The second quality assessment comes partly from the pipeline by measuring the resulting noise and comparing 
with the expected noise, but mainly from the astronomer who should visually inspect the pipeline products. The 
pipeline allows for inspection by providing products at all processing levels, and browse products at the final 
processing stage (Section \ref{sec:browse}). In Section \ref{sec:inspect}, the priority of inspecting beyond the 
browse products is suggested. 

\subsection{Browse products}
\label{sec:browse}
The final results of the HIFI pipeline processing were provided in two related products. The first was the stand-alone 
browse product itself. This product is the spectra or spectral cubes for each HIFI spectrometer of the observation as 
they are found at the end of the Level 2.5 processing.  

The second product is the browse image or postcard. This image offered a quick look at the data. It corresponded 
to the postcard displayed in the HSA User Interface (HUI) and showed the outcome of the Level 2.5 product 
generation. We note that these products were not necessarily science-ready. Although these postcard images gave a 
first impression of the quality of the data, often the browse images showed problems that needed to be 
addressed with further processing (e.g. baseline ripples). A high-resolution image of the browse product is 
available by downloading the {\it jpg} product (right-click on the postcard).

\begin{figure}[h]
\includegraphics[width=0.5\textwidth]{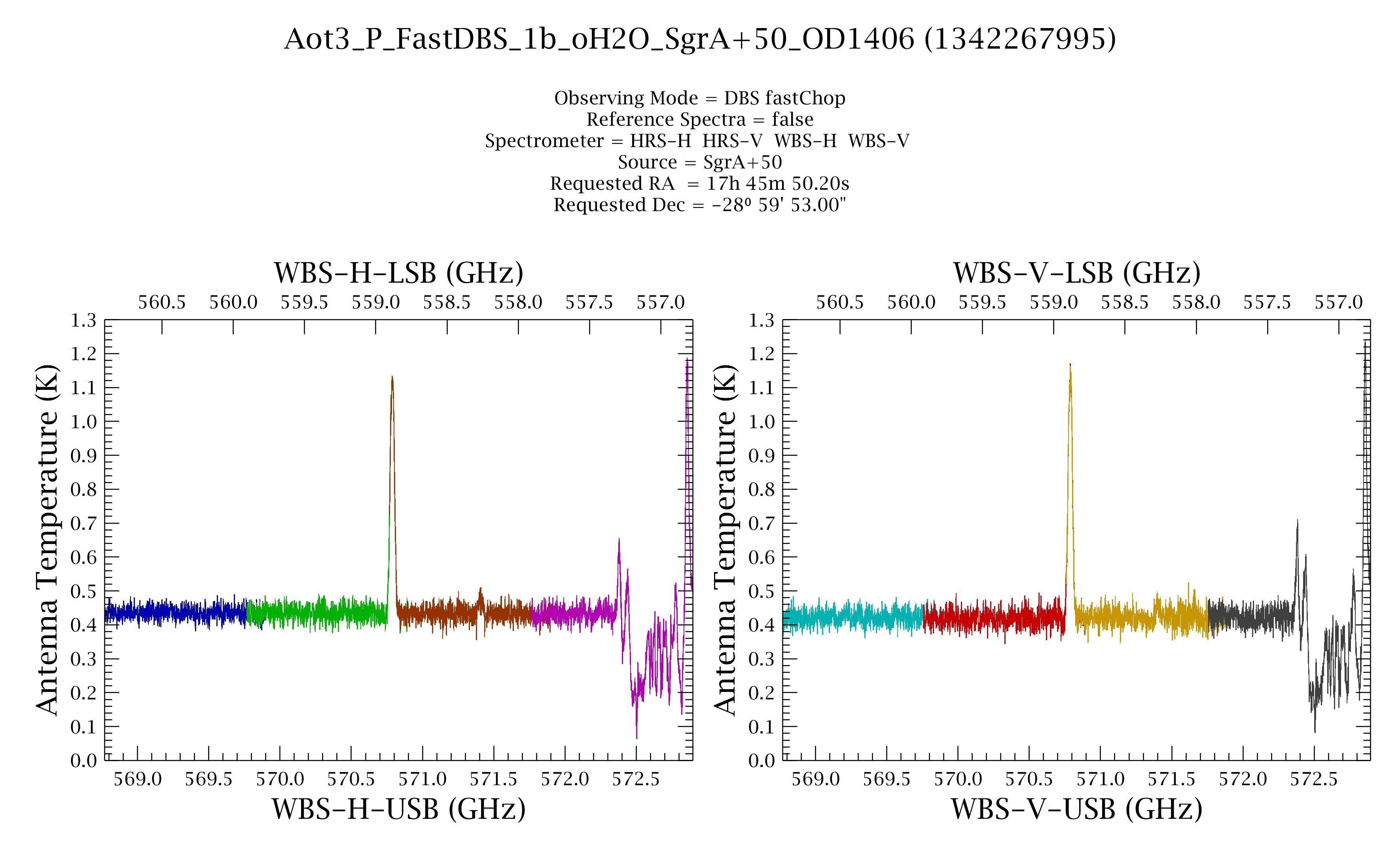}
\caption{Example of a browse image postcard provided with a HIFI DBS observation. The image is made from 
the observation WBS data, in each polarisation. The different colours are the WBS sub-bands. HRS data 
are not shown in the postcard.}
\label{fig:browse:point}
\end{figure}

For {\it Single Point mode}, the postcard shows some of the main observation parameters together with two 
plots of unstitched Level 2 WBS spectra with the H-polarisation to the left and the V-polarisation to the right. 
Each browse image lists the upper-sideband and the lower-sideband frequency scales. HRS spectra are 
not shown. Unless the spectral line is very strong, the images at first glance only show noise. The browse 
image does give a general impression whether the baseline in the data is flat, or not. Figure \ref{fig:browse:point} 
shows an example of a browse product image. Here, we note that the apparent absorption feature (572.3-572.8 
GHz - USB scale) is real { and due to the 557 water line in the lower sideband.} The Level 2.5 (see Section \ref{sec:levels:level25}) data are 
downloadable with the browse image for more careful inspection.

\begin{figure*}[!t]
\begin{center}
\includegraphics[width=0.95\textwidth]{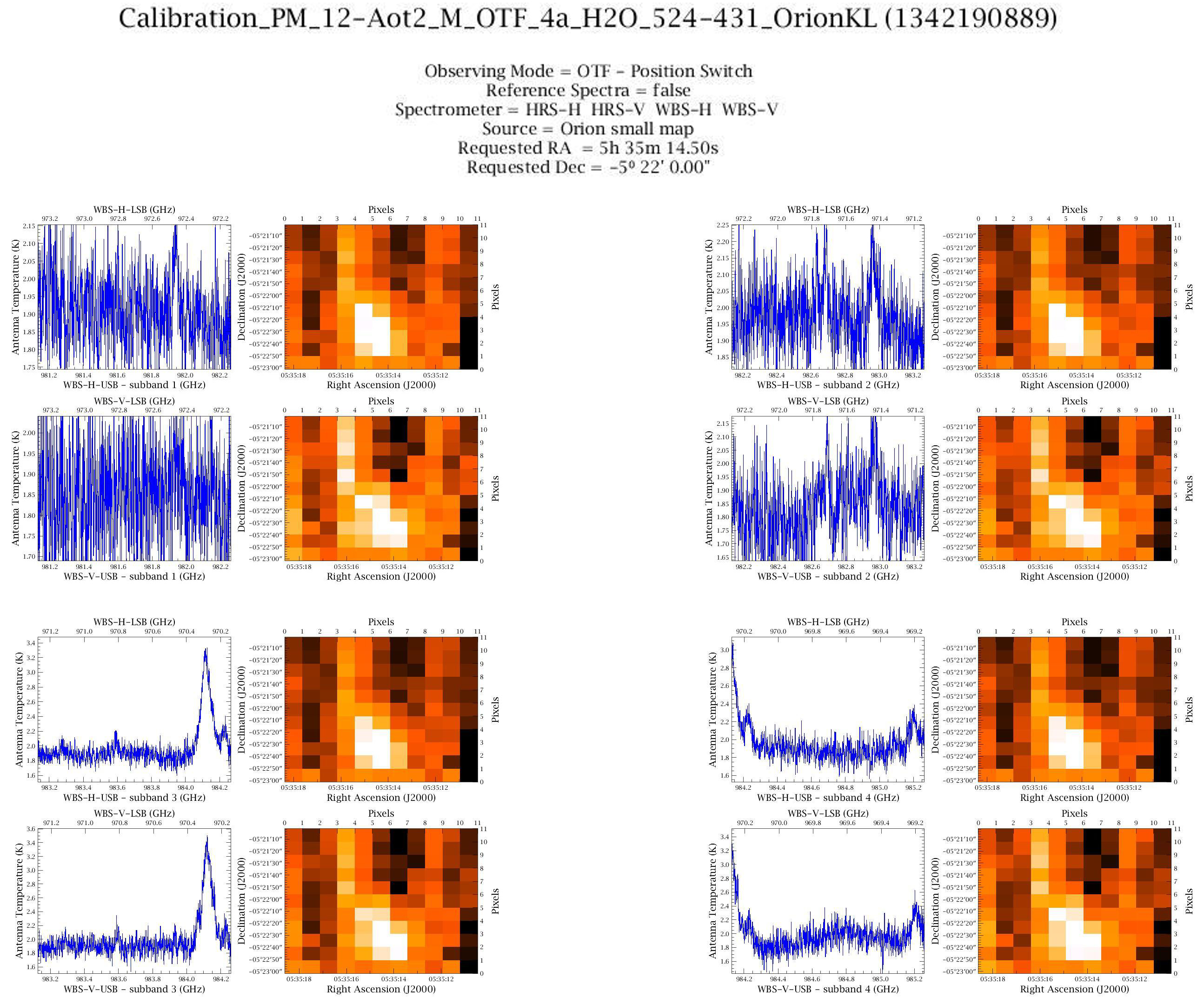}
\caption{Example browse images for a spectral cube. The image shows spectral cube data separately for each 
WBS sub-band and polarisation in the observation. The colour images are made by integrating over the 
frequencies in each sub-band, whereas the spectra are taken as an average over all positions. The top left panel shows 
WBS sub-band 1 H polarisation with V polarisation data just below. The top right panel shows sub-band 2, the bottom left panel shows
sub-band 3, and the bottom right panel shows sub-band 4. }
\label{fig:browse:cube}
\end{center}
\end{figure*}

For {\it Spectral Mapping mode}, the browse image shows sets of map-averaged Level 2 spectra for each 
sub-band along with the integrated map for that sub-band. For each sub-band, there are two sets of images, one
for each polarisation. The spectra are formed by averaging all spectra per map position while the integrated map 
is the full integration over the entire sub-band. The spectra and integrated maps were created without correcting  for any baseline problems (Section \ref{sec:additional}). When the baseline suffered from drifts and/or 
standing waves, the continuum dominates the map. The stand-alone browse data associated with the browse 
image are the spectral cubes with merged sub-bands per polarisation and spectrometer (see Section 
\ref{sec:levels:level25}). Figure \ref{fig:browse:cube} shows an example of the browse image for cubes. At the 
top of the images, the observation is described briefly.

\begin{figure*}[!t]
\begin{center}
\includegraphics[width=0.8\textwidth]{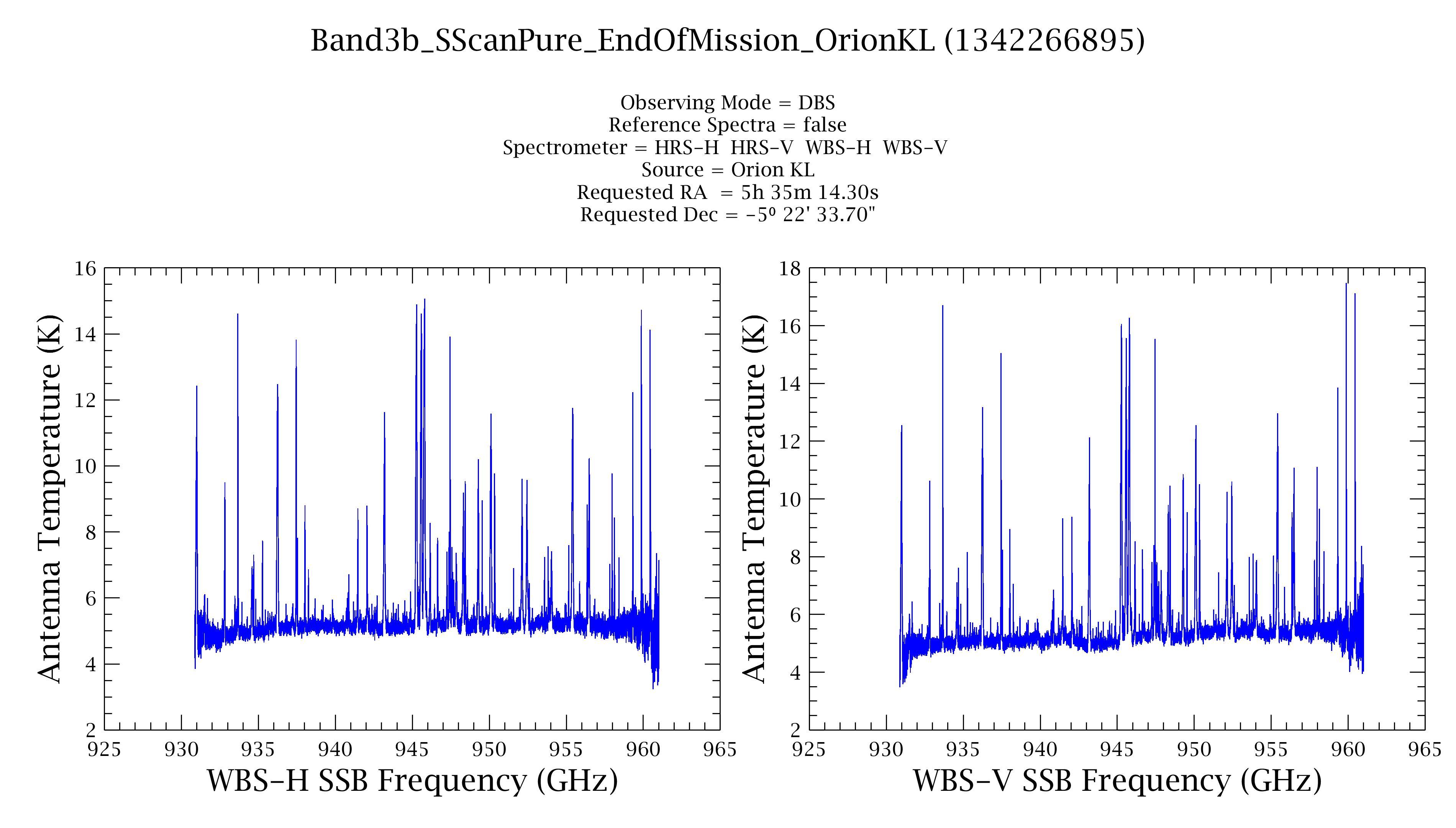}
\caption{Example browse images for a spectral scan. The data are the single-sideband spectra that resulted 
from the deconvolution. The two spectra show the H and V polarisations, respectively. The frequency range 
reflects the frequencies that are covered in the spectral scan.}
\label{fig:browse:sscan}
\end{center}
\end{figure*}

For {\it Spectral Scan mode}, the browse image shows the single-sideband solution after deconvolving  the 
Level 2 WBS spectra. No baseline correction was made before deconvolution. The H-polarisation is shown 
to the left and the V-polarisation to the right. The gap between the sidebands is shown as a line at 0 K. Figure 
\ref{fig:browse:sscan} shows an example of a browse image of a spectral scan. We note that the images are at 
significantly lower resolution than the information present in the data, and the images  only provide a rough indication of 
what the data can reveal.  

\subsection{Priority of inspecting pipeline products}
\label{sec:inspect}

Although the stand-alone browse products give a good indication of the overall data quality, for science 
purposes the co-added spectra that make up the browse products might wash out isolated artefacts or 
add data deemed unusable for science.  It is recommended that  individual spectra that make up the 
browse product be inspected before scientific analysis.
 
For Single Point mode observations, Level 2 spectra correspond to the co-addition of all individual spectra, 
and reflect problems that occurred throughout most of the integrations. Level 1 should be inspected only to 
identify problems with any specific integration. If any are found, it is recommended to work from Level 2 data before to 
the co-addition. This can be achieved by a particular reprocessing of the Level 1 data, but is also readily 
available via the HSA as {\it HIFI Non-averaged Level 2 Spectra} 
\footnote{\label{updp_fn} \href{http://herschel.esac.esa.int/twiki/pub/Public/HifiDocsEditableTable/HIFI_nonAveragedLevel2_UPDP_ReleaseNote.pdf} {HIFI\_nonAveragedLevel2\_UPDP\_ReleaseNote.pdf}} 
for all Single Point mode observations.
 
For the Spectral Mapping and Spectral Scan modes, inspecting each Level 2 spectrum should give a good 
idea as to whether standing waves or other instrument artefacts are still present in the final products. When 
this is the case, the data can be baseline corrected and will then need to be combined again into either 
spectral cubes (for mapping) or deconvolved spectra (for spectral scans).

Alternatively, a fraction of Spectral Mapping and Spectral Scan modes have been manually cleaned from 
baseline artefacts by instrument experts, and can be fetched as Highly Processed Data Products 
(footnote \ref{hpdp_fn}) from the Herschel Science Archive. 

\subsection{Quality reports}
\label{sec:qualityreport}
In automatic processing, most processing steps will have an expected outcome. When the outcome is 
noticeably different than expectations, the pipeline will raise a flag. Some of the differences are readily 
identified (e.g., a failure to reach a solution to a WBS comb measurement). Others, like comparing noise 
in an observation with respect to predicted noise, are more involved and require significant processing of 
the observational results. The resulting quality flags should then be taken as an indication of potential problem 
and not necessarily a true problem. In any event, it is worth the effort to examine the data. An overview 
quality report is provided by the HSA and in the Quality Control Summary Report in HIPE for each 
observation and provides an adequate starting guide for potential data problems.

Significant effort has been made to provide clear flag names and/or descriptions. At each level of processing 
quality items are reported. Chapter 10, Section 4 of the HIFI Data Reduction Guide (\HDRG see footnote 
\ref{hell_fn}) provides a complete listing of all quality flags and actions suggested to be taken. Each flag includes a description of the detected problems, an evaluation of the effect on science data, and a recommendation 
for how to proceed  in case that the flag is raised. The quality flags are separated into three levels of severity but these levels 
are not reflected in the Quality Summary Report provided to the user via the HSA. We note that the Quality Products 
provided with each level of data processing are the results of checks and are always populated with 
information even if nothing is found to be wrong whereas the Quality Report lists items that were flagged as 
problems.

\section{Additional processing}
\label{sec:additional}

For the majority of HIFI observations, Level 2 and Level 2.5 products can immediately be used for scientific 
analysis. However, as mentioned earlier, they were produced under the assumption that the referencing scheme 
sufficiently addressed all instabilities without additional correction for baseline drift or standing waves. 
Furthermore, the temperature scale or reference frame of the data may not be the most appropriate 
for a given scientific objective, such as combining HIFI data with other telescope data-sets.   

Within the interactive \texttt{HIPE} software package, the HIFI pipeline provides a mechanism to perform 
some additional steps. The 'Configurable Pipeline' feature of the \texttt{hifiPipeline} provides a way to 
rerun all or part of the pipeline in order to omit tasks or to change task defaults. Alternatively, it is possible 
to examine the results of each step of processing in order to improve data quality by accessing the ICC 
pipeline scripts used in the pipeline. It is also possible to use a 'non-ICC' pipeline algorithm that is based in part 
on the ICC algorithm. Section 5.4  of the \HDRG provides the necessary documentation on how to use  the interactive \texttt{hifiPipeline}.

The observation that is obtained from the HSA contained, along with the Level 0-2.5 data products, everything 
that is required to reprocess observations: calibration products, satellite data, as well as quality, logging, and 
history products which can be used to identify any problems with a data-set or its processing.

This section briefly describes the additional processing that may be required for a given observation. 
Section \ref{extra:fbase} addresses corrections to specific instabilities within HIFI data, while Sections 
\ref{extra:coaddhv}, \ref{extra:continuum}, and \ref{extra:k2jy} address more science-driven additional 
processing steps.

\subsection{Baseline distortions \label{extra:fbase} \label{extra:fhf}}
\begin{figure}[h]
\includegraphics[width=0.4\textwidth]{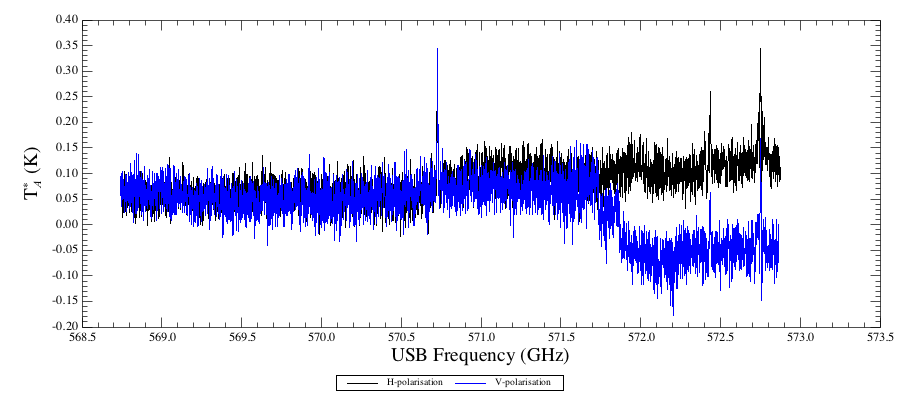}
\caption{The feature in the upper end of the V polarisation is an example of baseline distortions.  Such distortions can be corrected using the task \texttt{fitBaseline}. \label{fig:baseline}}
\end{figure}

\begin{figure}[h]
\includegraphics[width=0.4\textwidth]{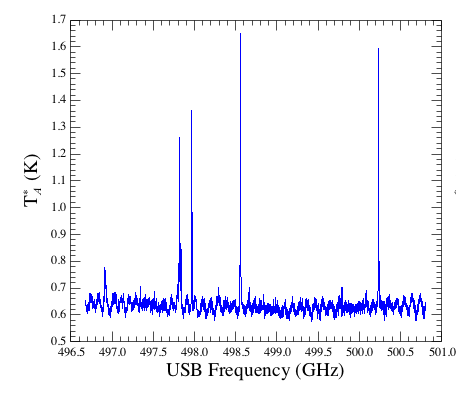}
\caption{Baseline ripples are an example of a standing wave in band 1a.  These ripples have a period near 100 MHz
due to the internal calibration loads. The ripples can be corrected for using the task \texttt{fitHifiFringe}. \label{fig:stw}}
\end{figure}

Common baseline distortions can take the form of platforming, residual standing waves, or other baseline 
instabilities: they are discussed with examples in Chapter 5.3.1 of the \HIFIHB. 
Within the HIPE software, 
there were two main tools to address these instabilities: \texttt{fitHifiFringe} for sine-fitting of standing
waves, or \texttt{fitBaseline} for fitting low-order polynomials to the baseline. Both of these modules are 
described in { more detail in} Chapter 12 and 13 of the \HDRG. { Figures \ref{fig:baseline} and \ref{fig:stw}
show two examples of baseline instabilities that are addressed by these modules} 

It is best to use the Level 2 spectra since they have not been sub-band stitched. When individual spectra are to be inspected before averaging at Level 2, the non-Averaged Level 2 UPDP described 
in footnote \ref{updp_fn} should be consulted.  This case applies mainly to Point Mode observations since neither maps nor 
spectral scans often contain many repeat spectra in any given Level 2 averaging.

\subsection{Co-adding H/V spectra \label{extra:coaddhv}}
Some processing steps were not taken by the standard pipeline. First, even though HIFI 
observed in two polarisations, the polarisations were not co-added. The pipeline did not average the polarisations for several reasons. It is possible that one polarisation was underpumped, or there 
were significant noise differences between the two polarisations. Another reason is that the polarisations 
were not perfectly spatially aligned and they could be seeing different features at slightly different positions 
(see Section 5.9 of  the \HIFIHB).    

However, when the data for the two polarisations are of sufficient quality and the small alignment differences 
are insignificant for a given object, averaging should be considered. The H and V point spectra and spectral 
scans can be co-added using the Level 2.5 spectra.  

With maps, this is best done before the initial spectral map is created using the Level 2.5 spectra (not the 
spectral cubes). These spectra should first be cleaned individually in each polarisation, then sent to the spectral -cube-building software (for example the task \texttt{doGridding} within HIPE or \texttt{map\_xy} within GILDAS/CLASS). 
The processing steps for the HIPE software are described in Chapter 15 of \HDRG ~ and in the GILDAS software 
package\footnote{\label{gildas_fn} \href{http://www.iram.fr/IRAMFR/GILDAS/}{http://www.iram.fr/IRAMFR/GILDAS/}}

The Highly-Processed Data Products available from the HSA of spectral maps in bands 6 and 7 have combined 
the polarisations and also inspected and removed most artefacts (see the footnote in Section \ref{sec:inspect}).  

\subsection{Correcting the continuum to a single-sideband scale 
\label{extra:continuum}} 

Owing to the double-sideband nature of the HIFI mixers, each Level 2 spectrum is provided in both the upper- and 
 lower-sideband frequency scales. This also implies that different sideband gain ratio calibrations are applied 
to the respective spectra. The Level 2 spectra are therefore provided with a single-sideband intensity scale, 
but this is only applicable to the information held in spectral lines since they only belong to one of the two 
sidebands. The continuum signal, on the other hand, is present in both sidebands and is therefore not properly 
calibrated in Level 2 spectra. Typically, it is a factor of two higher than the real single-sideband intensity. When 
the continuum needs to be accurately estimated (e.g. for absorption line studies), it is possible to recover the 
single-sideband continuum using the following recipe (see also Section 5.8.4 or the \HIFIHB):

\begin{equation} 
T^{\rm corr}_{\rm SSB}(\nu) = T_{\rm SSB}(\nu) - [C_{\rm DSB}-C_{\rm SSB}] = T_{\rm SSB}(\nu) - C_{\rm DSB} (1-G_{\rm ssb}) 
\label{eq:contabs} 
\end{equation} 

where $T_{\rm SSB}(\nu)$ is the Level 2 or 2.5 USB or LSB data, $C_{\rm DSB}$ is the estimated continuum from 
the Level 2 or 2.5 product, and $G_{\rm ssb}$ is the sideband gain applied to that sideband which can be found in 
the product header under the keyword {\tt usbGain} or {\tt lsbGain} depending on which sideband was used. We note 
that this correction assumes that the continuum emission does not differ between the two sidebands.  We also note 
that this is not a deconvolution since spectral features in both sidebands may be present in the resulting spectrum.

\subsection{Converting the data into $T_\textrm{MB}$ or Jy. \label{extra:k2jy}} 

The antenna temperature scale can also be easily changed into main-beam temperature T$_{\rm MB}$ or into Jansky. 
For the former, assuming a source filling the telescope main beam, the T$_{\rm MB}$ is simply given by:

\begin{equation} 	
T_{\rm MB} = {{\bf{\eta_l}} \over {\eta_{\rm MB}} } T_{\rm A}^* 
\end{equation} 

where {\bf $\eta_l$} is the forward efficiency (taken to be 0.96 for all HIFI bands), and $\eta_{\rm MB}$ is the main-beam efficiency for a source that fills the beam. This gain is found in the product header under \texttt{beamEff}. 

On the other hand, the conversion from \Tstar~ into Janskys requires making an assumption of the source 
size, as beam filling factors enter the equation \citep{HIFI-ICC-RP-2014-001}. Analytical formulae exist for perfectly 
Gaussian beams and simple source morphologies (e.g. Gaussian, top hat). A dedicated HIPE task, \texttt{convertK2Jy}, 
exists that allows performing this {conversion} with one of the simple source morphologies mentioned earlier, or even 
provide an ad hoc source brightness distribution. The task uses the full 2D model of the HIFI beams. The conversion 
factor for a perfectly { pointed} point-source is given in Table \ref{tbl:spotFreqs}. Further details on scale conversion can be 
found in Section 5.8.5 of the {\it HB}.

\begin{table}[tbp] 
\centering 
\begin{tabular}{rrrrrr} 
\hline \hline Mixer & $f$ & $\eta_{mb}$ & $\eta_A$ & HPBW & K-to-Jy \\ 
 & GHz&  &  & \arcsec & Jy/K \\ 
\hline 
1H & 480  & 0.62 & 0.65 & 43.1 & 482 \\ 
1V & 480  & 0.62 & 0.63 & 43.5 & 497 \\ 
2H & 640  & 0.64 & 0.64 & 32.9 & 489 \\ 
2V & 640  & 0.66 & 0.66 & 32.8 & 474 \\ 
3H & 800  & 0.62 & 0.63 & 26.3 & 497 \\ 
3V & 800  & 0.63 & 0.66 & 25.8 & 474 \\ 
4H & 960  & 0.63 & 0.64 & 21.9 & 489 \\ 
4V & 960  & 0.64 & 0.65 & 21.7 & 482 \\ 
5H & 1120 & 0.59 & 0.54 & 19.6 & 580 \\ 
5V & 1120 & 0.59 & 0.55 & 19.4 & 569 \\ 
6H & 1410 & 0.58 & 0.59 & 14.9 & 531 \\ 
6V & 1410 & 0.58 & 0.60 & 14.7 & 522 \\ 
7H & 1910 & 0.57 & 0.56 & 11.1 & 559 \\ 
7V & 1910 & 0.60 & 0.59 & 11.1 & 531 \\ 
\hline 
\end{tabular} 
\caption{Adopted values for $\eta_{mb}$, $\eta_A$, HPBW, and point-source sensitivity $S/\Tstar$ (i.e. Kelvin 
to Jansky conversion factor) $=(2k_B/A_g)(\eta_l/\eta_A)$ for one spot frequency per mixer. This table is to 
be compared to Table\,5 in \citet{2012A&A...537A..17R}, which uses the same spot frequencies, but outdated 
efficiencies and beam widths. \label{tbl:spotFreqs} } 
\end{table}

\section{Discussion}
\label{sec:discussion}

\subsection{Project lifespan}
Parts of the  HIFI pipeline were developed at a time well in advance of actual use. The Levels 0 and 0.5 
pipelines were in heavy use during instrument level testing before launch. The Level 2 pipeline, on the 
other hand, was initially a placeholder that averaged all Level 1 spectra. Keeping the steps well modularised 
helped the development, since any additional new step could be introduced into an existing pipeline script. In 
the end, this approach allowed the pipeline to mature as the understanding of the HIFI instrument and its data 
matured.  
 
Early development had the risk of introducing tacit and hard-to-correct assumptions about the instrument, satellite, 
calibrations, and the data products themselves. Perhaps, the most important early development was the definition 
of the data products. Data products reflect the instrument that produced them and the way in which the instrument 
was used. For HIFI, the main data features were instrument counts and { frequency } scales as functions of time. For 
HIFI, both the counts and the frequency scales could change with time. The implementation of the initial HIFI products 
was two-dimensional spectra. Data products also need to reflect the science uses, however. The final products must be 
usable by the community at large. Usability means either that the format of the product is standard enough to work with 
standard analysis software, or that there is available speciality software that understands the format of the new 
products.  

At the beginning of the project, it was tacitly assumed that the HCSS/HIPE user interface would continue well 
beyond post-operations. Since there was no plan for continued maintenance, the assumption went further to hope 
that eventually the HCSS/HIPE software would become standard analysis software. Going into post-operations, 
it looked doubtful that HIPE would continue to be supported beyond the end of the HSC. This meant that a better support of other data formats was required. For heterodyne spectroscopy, an exporter to the GILDAS/CLASS 
data format was developed mainly to support the HIFI consortium. During post-operations, Level 2.5 HIFI data products 
were introduced which were a simple reformatting of the PointSpectrum away from the HifiSpectrumDataset (2D 
spectra) to a more standard flux-vs.-frequency format (SimpleSpectrum). However, giving the restrictions of the 
HCSS/HIPE software, even this final format involved added complexity to a conceptually simple data product.   

Projects such as \Herschel must address not just the final archiving of the data products, but must also retain 
their utility. Usability can be achieved by defining  products that conform to standard formats (if they exist) or by 
maintaining specialist software over the long term and by providing complete documentation of the data formats 
used in the project. This point is further discussed below.

\subsection{\Herschel context}

The \Herschel ground segment made the decision to use common open source software for all instruments. 
Since no common software package existed that supported all three instruments, a new environment had to 
be created. This environment had to support  the instrument engineers, bulk data  processing, and the 
\Herschel user community, but also data products from the different instruments. The resulting software was 
the so-called Herschel Common Software System (HCSS) with HIPE providing the user interface.

The HCSS was a novel concept with far-reaching implications. The requirement was for open and common 
software for all \Herschel instruments. This further implied that the HCSS should operate on machines running 
under any "common" operating system and not use any proprietary software. Finally, the FITS table format was 
chosen to save and restore products.  

The requirement that all three \Herschel instruments use HCSS meant that all instruments were using software 
different than what each science community used  and indeed, since data product formats reflected both the 
instrument and the software, the HCSS products were often not directly usable in standard analysis software. 
To remedy this situation, the ground segment also took on the task of providing analysis software within the 
HCSS and a user environment.  { This extra effort required additional support from ESA in the form of software-development contracts.}


The HCSS/HIPE  provided the required environment for all the instrument teams to calibrate and pipeline 
process the data of their instrument. It also provided the necessary data management (tasks reading from the 
HSA) or inspecting contents of an observation, and many novel innovations required to support the diverse 
data products that \Herschel produces. In the end, HCSS/HIPE was able to process, display and manipulate images, 
spectra, and spectral cubes. Unfortunately, the user software (HIPE) was initially released  before it had 
reached sufficient maturity and user-friendliness as a scientific data analysis package, and, as a consequence, 
never gained wide user acceptance by the science community. 

The HIPE software will not be maintained beyond the end of \Herschel post-operations (although the final 
legacy version  will be provided via a virtual machine). One of the consequences of this is that many HIFI 
products that are easily accessible from within HIPE needed to be recast into stand-alone formats that 
can be accessed independent of HIPE from the HSA and/or the HSC user pages. The stand-alone products 
are noted in earlier sections as well as in the \HIFIHB. Currently, there are software packages other than HIPE 
to deal with HIFI products. Furthermore, all Herschel data products are in the FITS binary table format, which 
describes the data accurately and sufficiently ensuring that the HIFI products will remain accessible and 
interpretable for years to come.

\subsection{How well did the pipeline perform}

With the goal of removing instrument artefacts and preparing scientifically useful data, the HIFI pipeline 
systematically processed all HIFI observations to the highest processing level,  and by processing in modular 
steps, facilitated data analysis. Guaranteeing completely artefact-free { final} products, however, remained 
a challenge. Ground-based observatories rely heavily on the observers making the final call about their data. 
To achieve the science goals, the observer decides which data to clean and which data are filtered out. The 
final steps of ground-based processing are interactive.  

There are two fundamental problems with the final steps of pipeline processing. The first is that the pipeline can 
only make a reasonable estimate as to the overall science goals that resulted in the given observation. Thus 
maps, for example, are created with standard parameters that may not be the optimum for a given science 
question. Secondly, despite the stable \Herschel environment, the HIFI data still suffered from residual baseline 
artefacts that were due to imperfect bandpass calibration. These problems finally had to be addressed (interactively or otherwise)
outside of the pipeline through the intervention of instrument experts, and were made available to the community 
via the HSA as highly processed data products.

\section{Summary}
\label{sec:summary}

This article presented the concepts of the HIFI pipeline. The main concepts of the pipeline are listed below.
\begin{itemize}
	\item One pipeline for all HIFI observations that populates the Herschel Science Archive.
	\item Process HIFI data in levels of increasing uncertainty retaining results from all processing levels.
	\item Provide a pipeline that produces the same HIFI products either autonomously within the Herschel 
	Standard Product Generation environment or interactively in the Herschel Interactive Processing Environment.
	\item Assess processing quality for each processing step.
\end{itemize}

The HIFI pipeline has supported analysis of HIFI data since before the launch of \Herschel. The modular design was able to cope
with the many changes that faced the 15-year lifetime { and provided a smooth transition of the data processing from early instrument models through post-mission operations}. Along the way, calibrations were updated and new 
calibrations were introduced. Entirely new processing ideas were developed and employed. New processing 
levels were introduced allowing for significantly different product types to be supported.

The HIFI pipeline processed all HIFI observing requests and calibrations taken during the {\bf \Herschel} mission and performed 
initial quality assessments. The Herschel Legacy Science Archive is populated with the pipelined data and quality 
results. We described here are the processing steps that the products in the HSA underwent and the 
important quality measures taken. 

{ This article provides a context on how to understand the various data products and quality assessments provided with HIFI data as well as suggestions on how to inspect the HIFI data products themselves.  Along with presenting details on the pipeline levels, we also briefly described some of the limitations to automatic data-processing and additional processing steps beyond the pipeline.}


\begin{acknowledgements}
{HIFI has been designed and built by a consortium of institutes and university departments from across
Europe, Canada and the United States under the leadership of SRON Netherlands Institute for Space 
Research, Groningen, The Netherlands, and with major contributions from Germany, France and the US.
Consortium members are: Canada: CSA, University of Waterloo; France: CESR, LAB, LERMA, IRAM; 
Germany: KOSMA, MPIfR, MPS; Ireland, NUI Maynooth; Italy: ALI, IFSI-INAF, Osservatorio Astrofisico 
di Arcetri-INAF; Netherlands: SRON, TUD; Poland: CAMK, CBK; Spain: Observatorio Astron Nacional 
(IGN), Centro de Astrobiologia (CSIC-INTA). Sweden: Chalmers University of Technology - MC2, RSS \& 
GARD; Onsala Space Observatory; Swedish National Space Board, Stockholm University - Stockholm 
Observatory; Switzerland: ETH Zurich, FHNW; USA: Caltech, JPL, NHSC.}
\end{acknowledgements}

\bibliographystyle{aa}
\bibliography{hifi} 


\begin{appendix}
\section{Level 0 to 2.0 pipeline diagrams}
\label{sec:pipelinediag}

\begin{figure*}[h]
\begin{center}
  \includegraphics[width=0.70\textwidth]{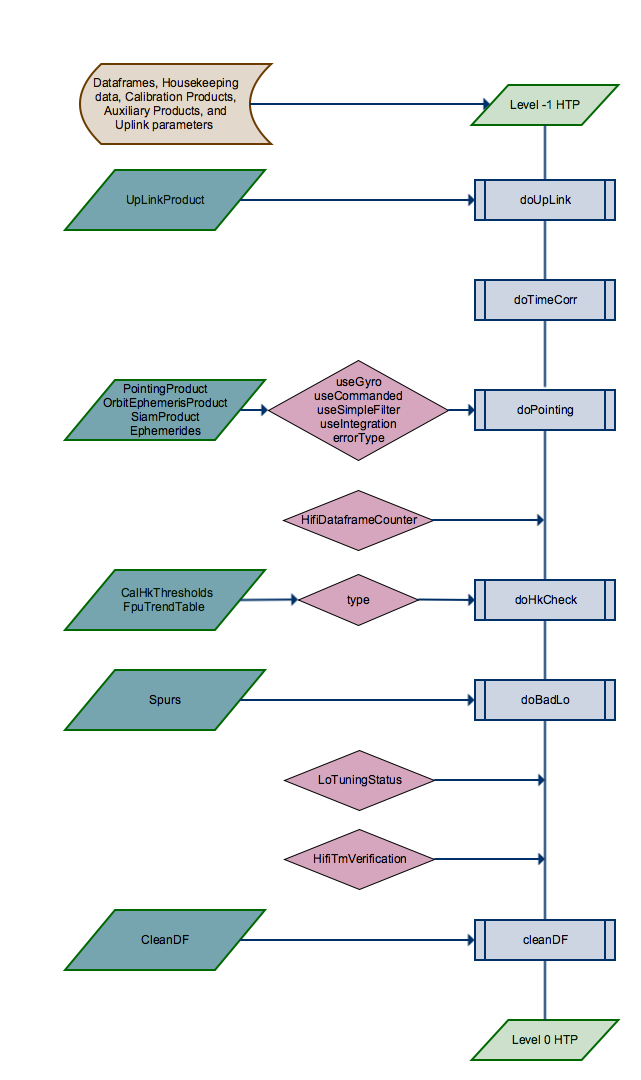}
  \caption{Level 0 pipeline diagram. The different shapes and colours have specific meanings. 
  Pipeline tasks are identified in light blue rectangles. The beginning and ending products are 
  shown as light green parallelograms. Calibration items are the dark green parallelograms. 
  Calibration can be either obtained from the calibration tree or created by a pipeline step. Task 
  options are identified within red rhombuses connected to a task. A red rhombus connected 
  directly to the pipeline indicates a quality-check step.}
  \label{fig:level0}
  \end{center}
\end{figure*}

\begin{figure*}[h]
\begin{center}
  \includegraphics[width=0.85\textwidth]{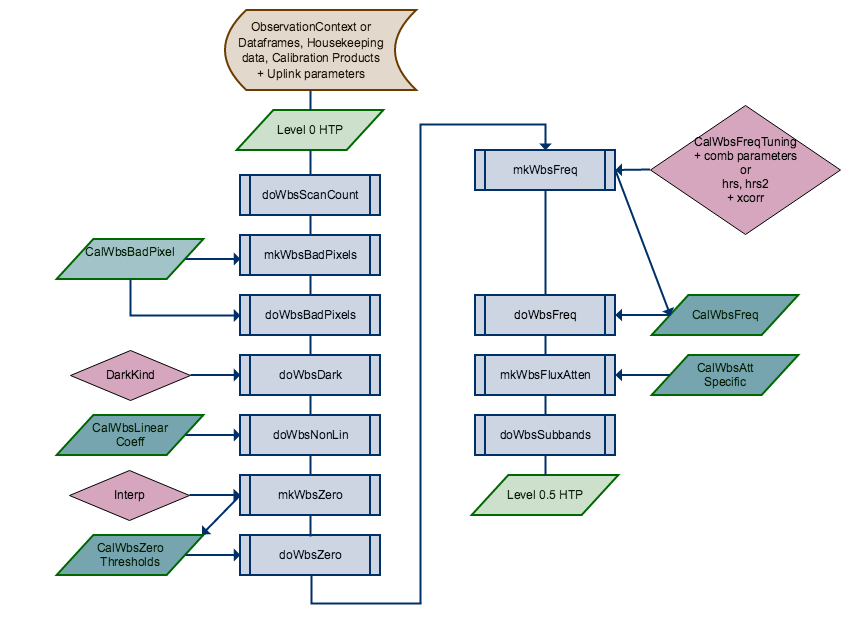}
  \caption{Level 0.5 pipeline diagram for the WBS. The input to this part of the pipeline is 
  Level 0 WBS data and calibration data in the {\it ObservationContext}. The symbols have 
  the same meaning as for the Level 0 pipeline in Figure \ref{fig:level0}.}
  \label{fig:level05wbs}
    \end{center}
\end{figure*}

\begin{figure*}[h]
\begin{center}
\includegraphics[width=0.85\textwidth]{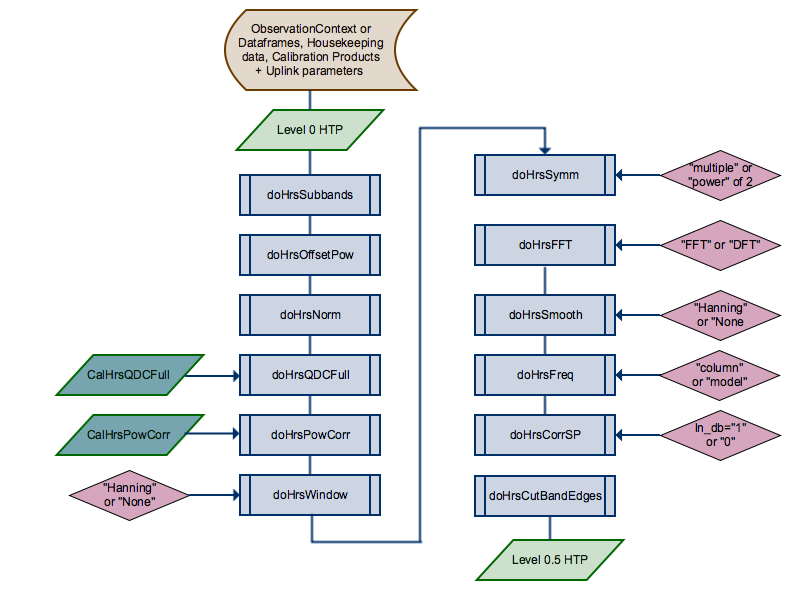}
    \caption{Level 0.5 pipeline diagram for the HRS. The input to this part of the pipeline is 
    Level 0 HRS data and calibration data in the {\it ObservationContext}. The symbols have 
    the same meaning as for the Level 0 pipeline in Figure \ref{fig:level0}.}
  \label{fig:level05hrs}
    \end{center}
\end{figure*}

\begin{figure*}[h]
\begin{center}
  \includegraphics[width=1.0\textwidth]{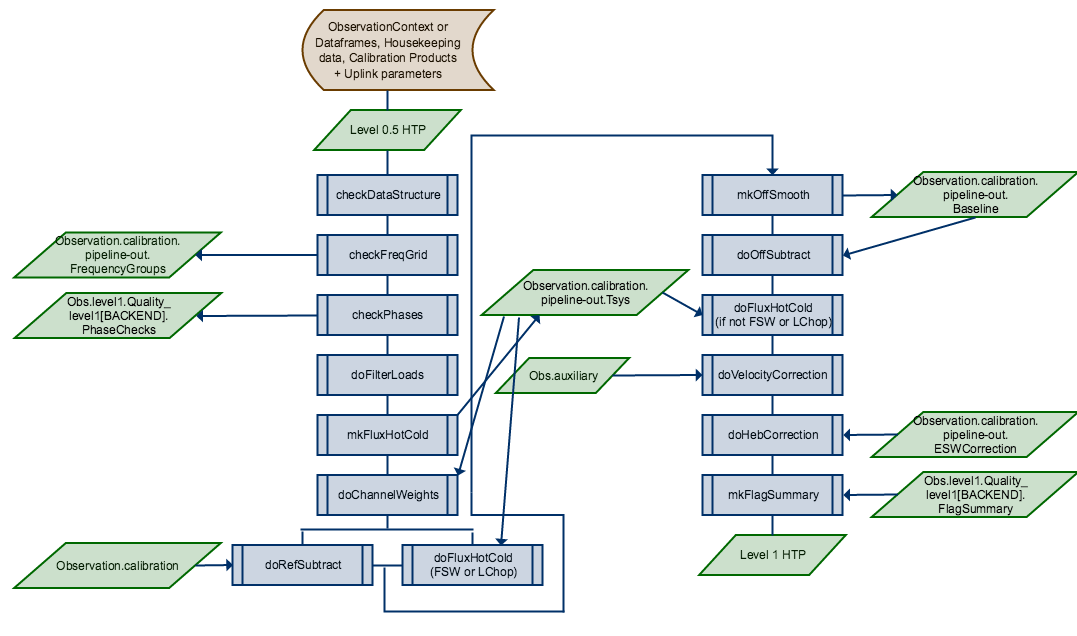}
  \caption{Level 1 pipeline flow diagram. The Level 1 pipeline uses the results of the Level 0.5 
  pipelines and processes the two backend in the same fashion. This and the Level 2 pipelines 
  are also often called the GenericPipeline. The Level 0.5 results were removed from the 
  {\it ObservationContext} to save disk space. The symbols and colours have the same meaning 
  as for the Level 0  pipeline in Figure \ref{fig:level0}.}
  \label{fig:level1}
    \end{center}
\end{figure*}

\begin{figure*}[h]
\begin{center}
  \includegraphics[width=0.9\textwidth]{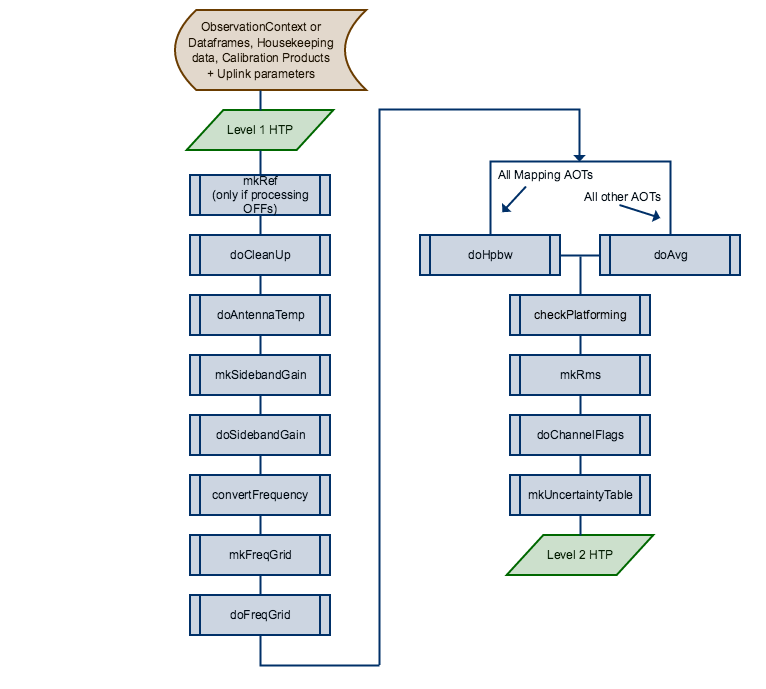}
  \caption{Level 2 pipeline diagram. The Level 2 pipeline uses the results of the Level 1 pipeline. 
  The symbols and colours in the figure have the same meaning as for the Level 0 pipeline in 
  Figure \ref{fig:level0}.}
  \label{fig:level2}
  \end{center}
\end{figure*}

From Level 0 to Level 2.0, various steps were taken in the pipeline. The following flow diagrams, 
{Figures \ref{fig:level0} to \ref{fig:level2}}, give a global view of the steps specific to each level. 
See Section \ref{sec:levels} for a brief description of each level.

The pipeline tasks are indicated in the light blue rectangles. Calibration products are shown in 
dark green.  {Task optional inputs are indicted with a red rhombus pointing to that task.  A red rhombus pointing to the pipeline flow indicates a quality-check step. }

\end{appendix}
\end{document}